\documentclass[preprint,authoryear, 12pt]{elsarticle}
\journal{International Journal of Forecasting}
\usepackage{cite}
\usepackage{natbib}
\usepackage[colorlinks=true, allcolors=blue]{hyperref}
\usepackage{amsthm}
\usepackage{amsmath, bm, amssymb}
\usepackage{graphicx}
\usepackage{subcaption}
\usepackage{adjustbox}
\usepackage{float}

\usepackage{threeparttable}
\usepackage{comment}

\usepackage{booktabs}
\newcommand{\bftab}{\fontseries{b}\selectfont}

\makeatletter
\def\ps@pprintTitle{%
 \let\@oddfoot\@empty
 \let\@evenfoot\@empty
}
\makeatother

\begin{document}

\begin{frontmatter}
\title{Beyond forecast leaderboards: Measuring individual model importance based on contribution to ensemble accuracy} 

\author[umass]{Minsu Kim\corref{cor1}}
\ead{tominsukim@gmail.com}
\cortext[cor1]{Corresponding author.}
\author[umass]{Evan L. Ray}
\ead{elray@umass.edu}
\author[umass]{Nicholas G. Reich} 
\ead{nick@umass.edu}

\affiliation[umass]{
    organization={School of Public Health and Health                  Sciences, University of Massachusetts},
    addressline={715 North Pleasant Street}, 
    city={Amherst},
    postcode={01003}, 
    state={Massachusetts},
    country={United States of America}}
    
\begin{abstract}
Ensemble forecasts often outperform forecasts from individual standalone models, and have been used to support decision-making and policy planning in various fields. As collaborative forecasting efforts to create effective ensembles grow, so does interest in understanding individual models' relative importance in the ensemble. To this end, we propose two practical methods that measure the difference between ensemble performance when a given model is or is not included in the ensemble: a leave-one-model-out algorithm and a leave-all-subsets-of-models-out algorithm, which is based on the Shapley value. We explore the relationship between these metrics, forecast accuracy, and the similarity of errors, both analytically and through simulations. We illustrate this measure of the value a component model adds to an ensemble in the presence of other models using US COVID-19 death probabilistic forecasts. This study offers valuable insight into individual models’ unique features within an ensemble, which standard accuracy metrics alone cannot reveal.
\end{abstract}

\end{frontmatter}

\section{Introduction} \label{sec:introduction}

Forecasting is a crucial challenge in various fields, including economics, finance, climate science, wind energy, and epidemiology. Accurate forecasts of future outcomes help individuals and organizations make informed decisions for better preparedness and more effective responses to uncertainties. Ensembles (or combinations) of individual forecasts are considered the gold standard because they generally provide more reliable performance in terms of accuracy and robustness than most, if not all, individual forecasts (\citet{Timmermann2006, Clemen1989, Gneiting2005, Viboud2018, Lutz2019}). 

In collaborative forecasting efforts, often the only forecasts that are used or communicated are those of the ensemble. 
For instance, during the COVID-19 pandemic, the US COVID-19 Forecast Hub (\url{https://covid19forecasthub.org/}) combined probabilistic forecasts from over 90 research groups to produce ensemble forecasts that retain the structure of a predictive distribution for cases, hospitalizations, and deaths in the US (\citet{Cramer2022-hubdata, Ray2023}). These ensemble forecasts were used by US Centers for Disease Control and Prevention (CDC) as official short-term forecasts to communicate with the general public and decision-makers (\citet{Cramer2022-evaluation}, \citet{cdc_web}). The Intergovernmental Panel on Climate Change (IPCC) also adopts a multi-model ensemble method to support the assessment of robustness and uncertainty arising from differences in model structures and process variability (\citet{RN10}) in its official reports, which policymakers use as a foundation for climate-related decisions and strategies.

In this work, we develop a measure of how much component models contribute to the skill of the ensemble. This model importance metric is generally applicable with a range of measures of forecast skill, for example, the squared prediction error (SPE) for point predictions and the weighted interval score (WIS) or log score for probabilistic forecasts (see details in \hyperref[subsec:accuracy-metric]{Section \ref{subsec:accuracy-metric}}). In the particular case of the expected SPE of a predictive mean, we show that our measure of model importance can be decomposed into terms measuring each component model’s forecast skill
as well as terms that can be interpreted as measuring how similar models’ predictions are to one another. Through simulated examples, we demonstrate that the insights from this decomposition in the point prediction setting can be extended to other contexts, such as scoring probabilistic forecasts using WIS.

\subsection{Motivating example}\label{subsec:motivating-example}

\begin{figure}[t!]
		\centering
		\includegraphics[width=\textwidth]{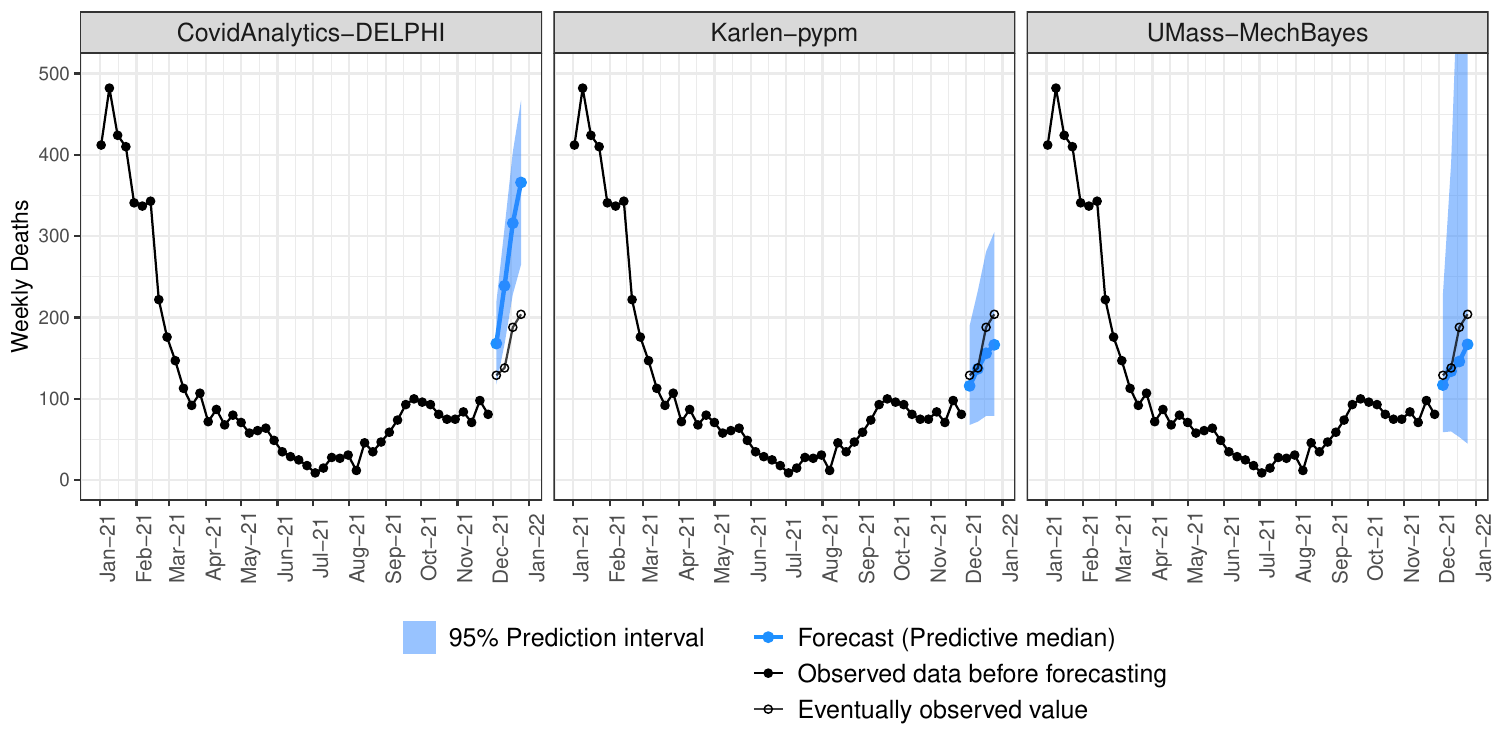}
		\caption{Distributional forecasts of COVID-19 incident deaths at 1- through 4-week horizons in Massachusetts made on November 27, 2021, by three models. Solid black dots show historical data available as of November 28. Blue dots indicate predictive medians, and the shaded bands represent 95\% prediction intervals. The open black circles are observations not available when the forecast was made. The 95\% prediction intervals of the UMass-MechBayes model (truncated here for better visibility of the observed data) extend up to 671 and 1110 for the 3-week and 4-week ahead horizons, respectively. }
		\label{fig:forecasts-example}
\end{figure}

The predictions from different models may differ depending on the model structure or input data sources used by the model. As an example, 
\hyperref[fig:forecasts-example]{Figure \ref{fig:forecasts-example}} shows the predictive distributions for incident deaths from three models submitted to the US COVID-19 Forecast Hub, along with eventually observed values. The quantile-based forecasts were made at 1-week through 4-week ahead horizons in Massachusetts on November 27, 2021. Here, two models under-predicted and one model over-predicted the outcomes.
The CovidAnalytics-DELPHI model has narrow 95\% prediction intervals and its forecasts are more biased than the other two models across all the four horizons, with especially large errors at forecast horizons of three and four weeks. On the other hand, the point estimates from the UMass-MechBayes model show less bias, but the predictive distributions are wide, especially broad for the 4-week ahead incident deaths. The forecasts of the Karlen-pypm model are moderate in both bias and prediction interval width.
These different predictive abilities are reflected in evaluation scores used for probabilistic forecasts. At the 4-week forecast horizon, the Karlen-pypm model had the best WIS with 20.4,  followed by UMass-MechBayes and CovidAnalytics-DELPHI with scores of 38.5 and 123.4, respectively. As discussed in more detail in \hyperref[subsec:case-study]{Section \ref{subsec:case-study}}, in this specific instance, the two models that were more accurate by traditional metrics actually proved to be less ‘important’ in the context of a multi-model ensemble because they were similar to multiple other submitted models.
In particular, those two models, along with most models that contributed to the ensemble, had a small downward bias, which was offset in part by over-predictions from the CovidAnalytics-DELPHI model. The predictions that were too high, while less accurate than those from other models, were the only ones that over-predicted the eventual outcome and therefore were important in offsetting a bias towards under-prediction from all the other models. This example motivates a closer examination of model importance within the ensemble forecasting framework.

\subsection{Related literature}\label{literature}

In the context of ensemble forecasting, some models will add more value than others. The problem of measuring each component model's impact on ensemble predictions bears methodological similarities to measuring variable importance in more standard regression-type modeling settings. Variable importance measures quantify the contribution of individual variables to the model's predictive performance. They are commonly used in model aggregation techniques such as Random Forest (\citet{Breiman2001}) and Gradient Boosting (\citet{Friedman2001}). 

As a related concept, Shapley values derived from a cooperative game theory measure how much each feature adds to the predicted outcome, on average, considering all possible coalitions of feature values. Building on the logic of the Shapley values, 
\citet{Giudici2021} proposed rank-based methodologies integrated with the Lorenz Zonoid approach to quantify the contribution of individual predictors in a regression-like setting. \citet{borup2024anatomy} also developed performance-based metrics tailored to the context of time series forecasting to achieve similar goals. 

Beyond evaluating the contribution of predictors, the concept of Shapley values has been employed to measure the value of component models in ensemble model settings. 
For example, \citet{Pompigna2018} applied this concept to a method for weighting component models in an ensemble model used to predict weekly/monthly fluctuations for average daily transport traffic. 
In the field of epidemic forecasting, \citet{Adiga2023} calculated Shapley-like values to determine the influence of each component model in the performance of a Bayesian Model Averaging ensemble at different forecasting phases of the COVID-19 pandemic. 

While these prior studies primarily focus on point predictions, our framework is designed for probabilistic forecasts, incorporating uncertainty quantification to capture the full distributional behavior of predictive models, which we identify as a central contribution of our work.

The value of including diverse models in an ensemble has been discussed in numerous studies. \citet{Batchelor-Dua1995} quantified diversity using the probability of a reduction in error variance, and the application of this metric to the data on US economic forecasts shows that ensembling is more beneficial when combining diverse forecasts than when combining similar forecasts.  
This idea has been supported by a number of follow-up studies, including \citet{Thomson2019, Lamberson-Page2012, Lichtendahl2020} and \citet{Kang2022}, which similarly emphasize that increasing the diversity of ensemble members improves overall ensemble performance. Furthermore, \citet{brown2005} have demonstrated the benefits of both reducing individual forecast error and maximizing diversity in individual forecasts. The ambiguity decomposition proved this to be a desirable feature of ensemble construction. This line of research provides a blueprint for achieving better accuracy in an ensemble: increasing the diversity of accurate individual forecasters. 
The model importance metric that we introduce, while related to the ambiguity decomposition, provides a more direct interpretation as it relates to ensemble forecast accuracy and it can serve as a general probabilistic forecast metric that can be used to evaluate, or even rank, the contributions of each component model to an ensemble.

\ \ 

The remainder of this paper is organized as follows. In \hyperref[sec:methods]{Section \ref{sec:methods}}, we address some accuracy metrics commonly used for point forecasts and quantile forecasts and introduce our proposed model importance metrics, including two algorithms for model importance metric calculation in the context of probabilistic forecasting. We also discuss the decomposition of the model importance metric in the context of point forecasts. 
We follow with some simulation studies to demonstrate that the insights from the decomposition based on point forecasting generalize to the probabilistic forecast setting and to examine the effect of bias and dispersion of a component model's distributional forecasts on the importance metric in the leave one model out algorithm setting.
\hyperref[sec:application]{Section \ref{sec:application}} provides results of applying the model importance metrics to real-world probabilistic forecast data from the US COVID-19 Forecast Hub. We present a case study investigating the relationship between the importance metric using the leave one model out algorithm and WIS with quantile-based forecasts of incident deaths in Massachusetts in 2021. Subsequently, we compare all the metrics to each other over a more extensive dataset. 
\hyperref[sec:discussion]{Section \ref{sec:discussion}} discusses the limitations of our study and outlines potential directions for further investigations. \hyperref[sec:conclusions]{Section \ref{sec:conclusions}}
concludes the paper with a summary of main findings and implications.

\section{Methods}\label{sec:methods}

\subsection{Accuracy metrics}\label{subsec:accuracy-metric}

Among the various forecast skill metrics developed to assess forecast quality, we focus on the mean squared prediction error for point predictions and the weighted interval score for probabilistic forecasts.

The squared prediction error (SPE) is defined as the square of the difference between the observed outcome $y$ and the predicted value $\hat{y}_i$ from model $i$:
\begin{equation}\label{PE}
    \text{SPE}(\hat{y}_i, y)=(y-\hat{y}_i)^2:=e_i^2.
\end{equation}
For the real-valued random variables $Y$ and $\hat{Y}_i$, the expected squared prediction error (ESPE) is formulated as 
\begin{equation}\label{EPE}
 \text{ESPE}(\hat{Y}_i, Y)= \mathbb{E}[(Y-\hat{Y}_i)^2],
\end{equation}
which quantifies the average squared discrepancy between the model's predicted values and the actual observed values.
ESPE accounts for the general performance of the model by considering the average error across all possible predictions and penalizes larger errors more significantly than smaller ones by squaring the differences between predicted and actual values. A lower ESPE indicates a model that makes predictions closer to the actual values on average.

The weighted interval score (WIS) of a probabilistic forecast is an approximation of commonly used probabilistic scoring rules such as the continuous ranked probability score (CRPS) and pinball loss. The WIS is expressed in terms of predictive quantiles as follows (\citet{Ray2023}):
\begin{align}\label{eq:wis}
    \text{WIS}({q{_{1:K}}},y)=\frac{1}{K}\sum_{k=1}^K 2\{\mathbf 1{_{(-\infty, q_{k}]}}(y)-\tau_k\}(q_k-y),
\end{align}
where $q_{1:K}$ denotes a set of $K$ distinct predictive quantiles, with $q_k$ representing the $k$th quantile evaluated at the quantile level $\tau_k$, for some positive integer $K$. For example, the predictive median corresponds to the quantile at $\tau_k=0.5.$ $y$ denotes the observed value and $\mathbf 1{_{(-\infty, q_{k}]}}(y)$ is an indicator function that equals 1 if $y \in (-\infty, q_{k}]$ and 0 otherwise.

The average of the accuracy metrics, either the SPE values or the WIS values, for model $i$ across many modeling tasks represents the overall performance of model $i$.

\subsection{Ensemble Methods}\label{subsec:ensemble-method}
Forecasts from multiple predictive models are aggregated to produce ensemble forecasts. 
The quantile-based forecasts are a common way to represent probabilistic forecasts, and
the predictive quantiles generated from each component forecaster are used for the quantile-based ensemble forecast. Let  different forecasters be indexed by $i$ $(i=1,2,\dots,n)$ and let $q_k^i$ denote the $k$th quantile from model $i$. The ensemble forecast value at each quantile level is calculated as a function of the component model quantiles
\begin{equation}\label{eq:quants}
    q_k^{\text{ens}}  = f(q_k^1,\cdots, q_k^n).
\end{equation}
 \hyperref[eq:quants]{Eq.(\ref{eq:quants})} is also applicable to point forecasts computing the ensemble prediction as a function of the point forecasts from component models. We note that the $q_k^i$ used hereafter refers to the $k$th quantile of a model $i$ for a specific forecasting task, which is a combination of the forecast location, date, and horizon. 

We employed the mean ensemble method, where all component forecasters have equal weight at every quantile level. 

\subsection{Model importance metric}\label{subsec:importance}
We propose two algorithms to evaluate a component model's contribution to an ensemble.
\subsubsection{Leave all subsets of models out (LASOMO)}\label{subsubsec:lasomo}

We utilize the Shapley value (\citet{Shapley1953}), a concept used in cooperative game theory. 

Let $N$ be the set of $n$ players in a game and $v$ be a real-valued characteristic function of the game. The {\bf Shapley value} $\phi_i$ of player $i \, (i=1,2,\dots, n)$ is defined as
\begin{equation} \label{eq:shapley}
	\phi_i = \sum_{\{S: S\subset N, \;i \notin S \}} \frac{s!(n-s-1)!}{n!}  \big[v(S \cup\{i\} )-v(S)\big],
\end{equation}
where $S$ is a coalition that consists of $s$ players out of the total of 
$n$ players, excluding player $i$ ($s\in \{0,1,2,\dots, n-1\}$). When $s = 0$, it indicates that $S=\emptyset.$

The characteristic function $v$ is assumed to satisfy $v(\emptyset)=0$ and for each subset $S$, $v(S)$ represents the gain
that the coalition can achieve in the game. Accordingly $v(S \cup\{i\} )-v(S)$ in \hyperref[eq:shapley]{Eq.(\ref{eq:shapley})} represents the marginal contribution of player $i$ to the coalition $S$ and its weight is computed by considering all possible permutations of players in $S$. An interpretation of this concept can be found in Section 1 of the supplement for further reference.

We calculate the importance metric of a component model in ensemble creation using this Shapley value. 
The $n$ players and a coalition of $s$ players in the game correspond respectively to the $n$ individual forecasting models and a collection of $s$ component models for an ensemble in our context.  A proper scoring rule serves as the characteristic function. However, this choice of the characteristic function does not satisfy the assumption that the value of the empty set is zero, as any scoring metric cannot be applied to an empty set of forecasting models, which means no prediction. It is also not meaningful to assign a quantitative score to ``no prediction''. To avoid this difficulty, we modify \hyperref[eq:shapley]{Eq.(\ref{eq:shapley})} to eliminate the case of the empty subset, and consequently, the denominator in \hyperref[eq:shapley]{Eq.(\ref{eq:shapley})} is replaced with $(n-1)!(n-1)$. (See also Section 1 of the supplement).

For a single forecast task $\tau$, the importance metric (i.e., the contribution) of the component model $i$ is calculated by
\begin{equation}\label{eq:imp_score}
    \begin{aligned}
    {\phi_{i}}_{\tau} = \sum_{\{S: S\subset N, \;i \notin S \}} \frac{s!(n-s-1)!}{(n-1)!(n-1)}  \Big[&\mu(F_{\tau}^{S \cup\{i\}}, y_{\tau})-\mu(F_{\tau}^{S}, y_{\tau} )\Big],
    \end{aligned}
\end{equation}
where $F_{\tau}^{A}$ represents the ensemble forecast constructed based on the forecasts from models in the set $A$, $y_{\tau}$ denotes the actual observation, and $\mu$ represents a positively oriented proper scoring rule. The difference in $\mu$ reflects the extent to which the component models contribute to the accuracy of the ensemble. A positive value of ${\phi_{i}}_{\tau}$ indicates that, on average across all coalitions of models, including the $i$th forecaster in the ensemble construction produces improved ensemble predictions. On the other hand, a negative value of ${\phi_{i}}_{\tau}$ means that including the $i$th forecaster in ensemble construction degrades ensemble prediction performance on average. 

The average of importance metrics of model $i$, ${\phi_{i}}_{\tau}$'s, across all tasks is the overall model importance metric of the model $i$:
\begin{equation}\label{eq:overall_imp_score}
    \Phi(i) = \frac{1}{|{\cal T}|} \sum_{\tau \in {\cal T}} {\phi_{i}}_{\tau}, 
\end{equation}
where ${\cal T}$ represents a collection of all possible forecasting tasks, and $|{\cal T}|$ indicates its cardinality.

We note that the weight for a subset in \hyperref[eq:imp_score]{Eq.(\ref{eq:imp_score})} is calculated in the same manner as in the Shapley value formula in \hyperref[eq:shapley]{Eq.(\ref{eq:shapley})} that is the weighted average over all possible permutations of coalitions. In this formulation, the weight depends on the subset size. The rationale behind this approach is that as more models are involved, the marginal contribution of an additional model tends to decrease, since new models are more likely to provide redundant information in the ensemble. However, alternative weighting schemes are possible. 
For example, \citet{Adiga2023} used equal weights to all subsets regardless of their size, meaning that the importance metric is an evenly weighted average of the marginal contribution over the subsets.

\subsubsection{Leave one model out (LOMO)}\label{subsubsec:lomo}

In addition to the Shapley value analysis using ``all subsets", we measure the ensemble forecast performance when a single model is removed from the ensemble in order to see how much that component model contributes to improving the ensemble accuracy. 
Let $S^{-i}$ denote the set of all models excluding model $i$, i.e., $S^{-i}=\{1,...,n\}\setminus \{i\},$ where $i=1,2,...,n$. Then, $F^{S^{-i}}$ represents the ensemble forecast built based on the forecasts from models in the set $S^{-i}$. That is, we remove the $i$th forecaster from the entire set of $n$ individual forecasters and create an ensemble from the rest. Similarly, $F^{S^{-i} \cup\{i\}}$ represents the forecast from an ensemble model that includes all $n$ individual forecasters.
The importance metric of the component forecaster $i$ for a single task $\tau$ is measured by
\begin{equation}\label{eq:imp_score_lomo}
    \begin{aligned}
    {\phi_{i}}_{\tau} = \mu(F_{\tau}^{S^{-i} \cup\{i\}}, y_{\tau})-\mu(F_{\tau}^{S^{-i}}, y_{\tau} ).
    \end{aligned}
\end{equation}

\subsection{Decomposition of importance metric measured by the LOMO algorithm based on point predictions}\label{subsec:decomposition}

In this section, we discuss components of importance metrics measured by the LOMO algorithm in the context of point predictions and their mean ensemble, which serve as a tractable starting point before extending to probabilistic forecasts. 

We use the positively oriented squared prediction error ($-\text{SPE}$) to assess the accuracy of the predicted values, since $-\text{SPE}$ facilitates a more intuitive interpretation of the resulting importance metric: a positive score indicates a beneficial impact on ensemble accuracy, while a negative score reflects a detrimental impact.

For $n$ point forecasts $\hat{y}_1, \hat{y}_2, ..., \hat{y}_n$ and the actual outcome $y$, the importance metric of model $i$ is calculated by subtracting the negative SPE of the ensemble forecast made from the predictions of all models except model $i$ from that of the ensemble forecast based on predictions from all $n$ models, written as
\begin{align} 
    \phi_i &= -\left(y-\frac{1}{n}\sum_{j=1}^{n} \hat{y}_j\right)^2 + \left(y-\frac{1}{(n-1)}\sum_{j\ne i} \hat{y}_j\right)^2 \label{eq:pt-pred-phii1}\\
    & = -\frac{1}{n^2} e_i^2 -\frac{2}{n^2}\sum_{j\ne i} e_ie_j + \frac{2n-1}{[n(n-1)]^2} \left(\sum_{j\ne i} e_j^2 + 2\sum_{\substack{j\ne i\\ j<k}} e_je_k\right), \label{eq:pt-pred-phii2}
\end{align}
where $e_j$ indicates the prediction error between $y$ and the predicted value $\hat{y}_j$ from model $j$ $ (j=1,2,...,n)$. Details of the process leading to \hyperref[eq:pt-pred-phii2]{Eq.(\ref{eq:pt-pred-phii2})} from \hyperref[eq:pt-pred-phii1]{Eq.(\ref{eq:pt-pred-phii1})} are available in the supplementary materials (see Supplemental Section 2).

The expected score is given as
\begin{align} \label{eq:mean-phii}
    \begin{split}
        \mathbb{E}(\phi_i) =& -\frac{1}{n^2}\text{ESPE}(\hat{Y}_i)+\frac{2n-1}{[n(n-1)]^2}\sum_{j\ne i} \text{ESPE}(\hat{Y}_j) \\
        &-\frac{2}{n^2}\sum_{j\ne i} \mathbb{E}(e_ie_j) + \frac{2(2n-1)}{[n(n-1)]^2}\sum_{\substack{j\ne i\\ j<k}}\mathbb{E}( e_je_k).
     \end{split}
\end{align}
The expected importance metric of model $i$ consists of two kinds of terms. The ESPE terms capture the accuracy of the individual models. The first term shows that the expected importance of model $i$ is lower when that model has a large ESPE, while the second term shows it is lower when the combined ESPE of the other models is small. 
The terms involving the product of prediction errors examine how two models' predictions relate to each other and to the actual observation. If the product of their errors is negative, it means those models' predictions are on opposite sides of the actual value (one overestimating while the other underestimates). The third term measures how much model $i$ helps to correct the errors made by the other models; as the combined expected correction increases, the expected importance of model $i$ increases. The last term indicates that when the forecast errors of other models are highly similar to each other, model $i$ is expected to be more important. 

It is worth noting that while our decomposition is closely related to that from \citet{brown2005} (see details in Section 2.1 of the supplement), our decomposition directly reveals that model $i$ is rewarded if it is not correlated with others and if the other models are correlated with each other.
We note that under the assumption of unbiased forecasts, the expected product of the errors corresponds to the covariance of the errors.

\subsection{Simulation Studies}\label{subsec:simulation}

In these simulation studies, we show that the decomposition insights developed in the point forecast setting remain applicable to probabilistic forecasts. We then explore the effect of bias and dispersion of a component model's predictive distribution on the importance of that model using the mean ensemble method in the LOMO algorithm setting. We focused on LOMO in these experiments because it closely aligns with the theoretical framework in \hyperref[subsec:decomposition]{Section \ref{subsec:decomposition}} and is simple to interpret. 

We created a set of three simulation scenarios to study model importance. The first two scenarios investigate model importance working with component point forecasts and probabilistic forecasts that have varying degrees of bias (\hyperref[subsubsec:simu-settingA]{Section \ref{subsubsec:simu-settingA}}). The third scenario investigates model importance with probabilistic forecasts with misspecified dispersion (\hyperref[subsubsec:simu-settingB]{Section \ref{subsubsec:simu-settingB}}). We assume that the truth values follow the standard normal distribution 
$$ Y_\tau \sim N(0, 1), \quad \text{for all }\tau \in \cal T,$$ 
where $\mathcal{T} = \{1, \dots, 1000\}.$ For each of the probabilistic scenarios we use 23 quantiles to represent the forecast distributions at the same quantile levels as in the data set used in the applications (see \hyperref[subsec:data]{Data}). We calculate the importance metric for each model based on individual observations, using the negative SPE for point forecasts and the negative WIS for quantile forecasts. As mentioned in \hyperref[subsec:decomposition]{Section \ref{subsec:decomposition}}, we adopt a positive orientation for a more straightforward interpretation, so that larger values reflect a more beneficial effect on the ensemble accuracy. The overall model importance is then taken as the average over ${|\cal T|}=1000$ replicates, which is an approximation to the expected importance metric.

\subsubsection{Setting A: Relationship between a component forecaster's bias and importance}\label{subsubsec:simu-settingA}

\begin{figure}[t!]
    \centering
    \includegraphics[width=\textwidth]{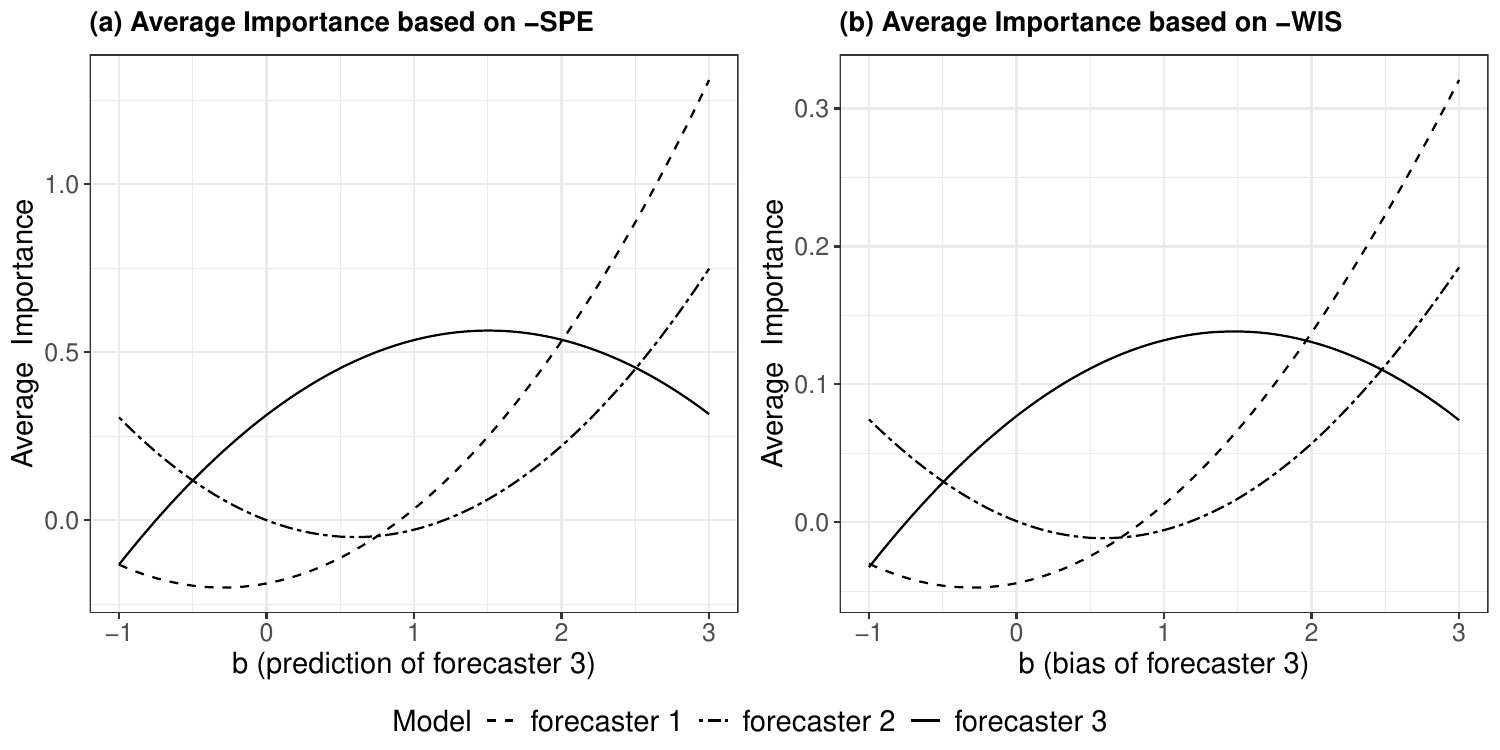}
    \caption{Expected importance of three forecasters as a function of the prediction/bias of forecaster 3 in simulation settings: (a) $\hat{y}_{1} = -1$, $\hat{y}_{2} =-0.5$, and $\hat{y}_{3} =b$ based on the negative SPE, (b) $F_{1,\tau} = N(-1, 1)$, $F_{2,\tau} = N(-0.5, 1)$, and $F_{3,\tau} = N(b, 1)$ based on the negative WIS, where $\tau=1,\dots,1000$. The data generating process is $N(0,1).$ The expected importance metrics were calculated and averaged over $1000$ replicates of the forecasting experiments conducted at each value of $b$, incremented by 0.05 from $-1$ to $3$.}
    \label{fig:simuplot-settingA}
\end{figure}

For the first scenario, we consider the following three point forecasts: 
\begin{align}\label{simu-a1}
	\hat{y}_{1} = -1,\quad
	\hat{y}_{2} = -0.5,\quad
	\hat{y}_{3} = b,
\end{align}
and in the second scenario, we assume that all the three component forecasters produce normally distributed forecasts as follows:
\begin{align}\label{simu-a2}
	F_{1,\tau} = N(-1, 1) ,\quad
	F_{2,\tau} = N(-0.5, 1) ,\quad
	F_{3,\tau} = N(b, 1),
\end{align}
where $\tau$ denotes the index of a generic replicate, with $\tau = 1, \dots, 1000$.
With the probabilistic component forecasts in \hyperref[simu-a2]{Eq.(\ref{simu-a2})}, the ensemble forecast distribution is $\displaystyle F_{\tau} = N\left({(b-1.5)}/{3}, 1\right )$. Note that the ensemble prediction is unbiased when $b = 1.5.$
We changed the value of $b$ from $-1$ to $3$ in increments of $0.05$ to observe how the importance of model 3 changes in both scenarios.
Note that the point predictions correspond to the means of the probabilistic forecasters.

The simulation results show that the importance metric of each forecaster matches the calculation derived in \hyperref[eq:mean-phii]{Eq.(\ref{eq:mean-phii})} (\hyperref[fig:simuplot-settingA]{Figure \ref{fig:simuplot-settingA}(a)}). Additionally, the general patterns of importance metrics observed for the three probabilistic forecasters closely align with the patterns seen with the point forecasters (\hyperref[fig:simuplot-settingA]{Figure \ref{fig:simuplot-settingA}(b)}).

In both settings, the forecaster that produces the least biased forecast achieves the highest importance metric when all the three forecasters have negative biases (i.e., biases in the same direction). However, when forecaster 3 has a small positive bias unlike the other forecasters, it becomes the most valuable component model in the accurate ensemble creation, as it serves to correct the negative bias of the other component models. 
If forecaster 3 has a large bias ($b\ge 2$), then, although it is the only model biased in the opposite direction, forecaster 1 becomes the most important contributor to the ensemble. This is because forecaster 1 plays a more considerable role in offsetting that large bias compared to forecaster 2.

\subsubsection{Setting B: Relationship between component forecaster dispersion and importance}\label{subsubsec:simu-settingB}

In this simulation scenario, there are three probabilistic forecasts, each equal to a normal distribution with mean $0$ and a different standard deviation:
\begin{align*}
	F_{1,\tau} = N(0, 0.5^2),\quad
	F_{2,\tau} = N(0, 0.7^2),\quad
	F_{3,\tau} = N(0, s^2),
\end{align*}
where $\tau$ denotes the index of a generic replicate, with $\tau = 1, \dots, 1000$. 
In this setup, both forecaster 1 and 2 have predictive probability distributions that are underdispersed compared to the distribution of the data generating process, which is $N(0,1^2)$. With these component forecasts, the standard deviation of the ensemble forecast distribution is calculated as $(0.5+ 0.7+ s)/3$ (see details in Section 3 of the supplement). Note that the ensemble is correctly specified when $s = 1.8.$
We changed $s$, the standard deviation of the forecast distribution produced by forecaster 3, from $0.1$ to $3$ in increments of $0.05$. 

\begin{figure}[t!]
    \centering
    \includegraphics[width=\textwidth]{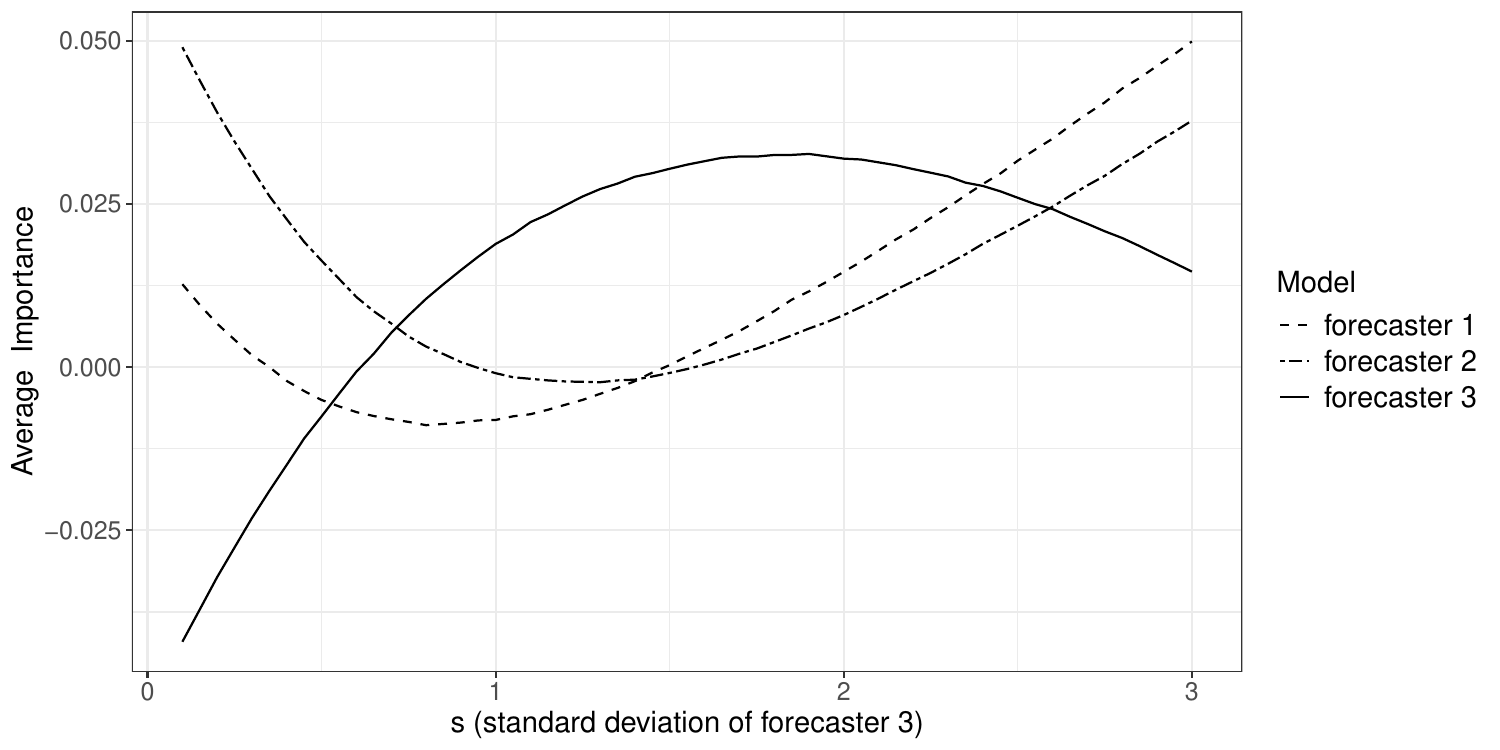}
    \caption{Expected importance of three forecasters as a function of dispersion of forecaster 3 in the simulation setting: $F_{1,\tau} = N(0, 0.5^2)$, $F_{2,\tau} = N(0, 0.7^2)$, and $F_{3,\tau} = N(0, s^2)$ based on the negative WIS, where $\tau=1,\dots,1000$. The data generating process is $N(0,1).$ The expected importance metrics were calculated and averaged over $1000$ replicates of the forecasting experiments conducted at each value of $s$, incremented by 0.05 from $0.1$ to $3$.}
    \label{fig:simuplot-settingB}
\end{figure}

\hyperref[fig:simuplot-settingB]{Figure \ref{fig:simuplot-settingB}} plots the expected or average importance metrics for the three forecasters as a function of the value of $s$. 
If the standard deviation of forecaster 3's predictive probability distribution is less than or equal to $0.5$ (i.e., $s\le 0.5$), then including that forecaster in the ensemble construction makes the ensemble's probabilistic forecast distribution narrower than not including that forecaster. This would make the ensemble's prediction very different from the truth, resulting in the forecaster having the lowest importance metric among all models. Starting from $s= 0.7$, forecaster 3 becomes the most important model, as the standard deviation of the ensemble’s forecast distribution with that model included approaches that of the truth distribution more closely than the ensemble without it, as $s$ increases.
For $s\ge1$, the predictions of $F_3$ become more and more overdispersed as $s$ grows, and this large variance brings the dispersion of the ensemble close to the truth; however, beyond a certain point, the ensemble predictions become more dispersed than the truth. Thus, forecaster 3 maintains its top ranking in the importance until $s$ reaches approximately $2.4.$
Thereafter, the ensemble formed without forecaster 1 generates forecasts with high dispersion, which results in the forecaster 1 having the highest importance metric among all models.

\section{Application}\label{sec:application}

In this application, we used probabilistic forecasts of COVID-19 deaths in the United States to evaluate each component model's contribution to probabilistic ensemble forecasts produced by the mean ensemble method.
In a case study (\hyperref[subsec:case-study]{Section \ref{subsec:case-study}}), we utilized the LOMO measure to provide a clearer illustration of the key intuitive insights, as LOMO offers more straightforward interpretability compared to LASOMO. A more extensive application is presented in \hyperref[subsec:extensive-application]{Section \ref{subsec:extensive-application}}, where both LASOMO and LOMO algorithms were applied and compared. \\

The code used for loading data and conducting all analyses and simulations is archived on Zenodo\footnote{\url{https://doi.org/10.5281/zenodo.17954018}} for reproducibility. The latest version of the associated code and data are available on GitHub\footnote{\url{https://github.com/mkim425/replication_model-importance}}.

\subsection{Data}\label{subsec:data}

The forecast data employed in this analysis were obtained from the US COVID-19 Forecast Hub that collected short-term quantile forecasts on COVID-19 deaths from various models developed by academic, industry, and independent research groups, from its launch in April 2020 (\citet{Cramer2022-hubdata}) through April 2024. The submitted forecasts were provided using 23 quantiles (0.01, 0.025, 0.05, 0.10, 0.15, . . . , 0.90, 0.95, 0.975, 0.99). The death data on COVID-19 from Johns Hopkins University Center for Systems Science and Engineering (JHU-CSSE) were used as the ground truth data (\citet{Dong2020}).

\subsection{Case study: Relationship between importance metric and WIS with data for deaths in Massachusetts in 2021}\label{subsec:case-study}

\begin{figure}[t!]
    \includegraphics[width=\textwidth]{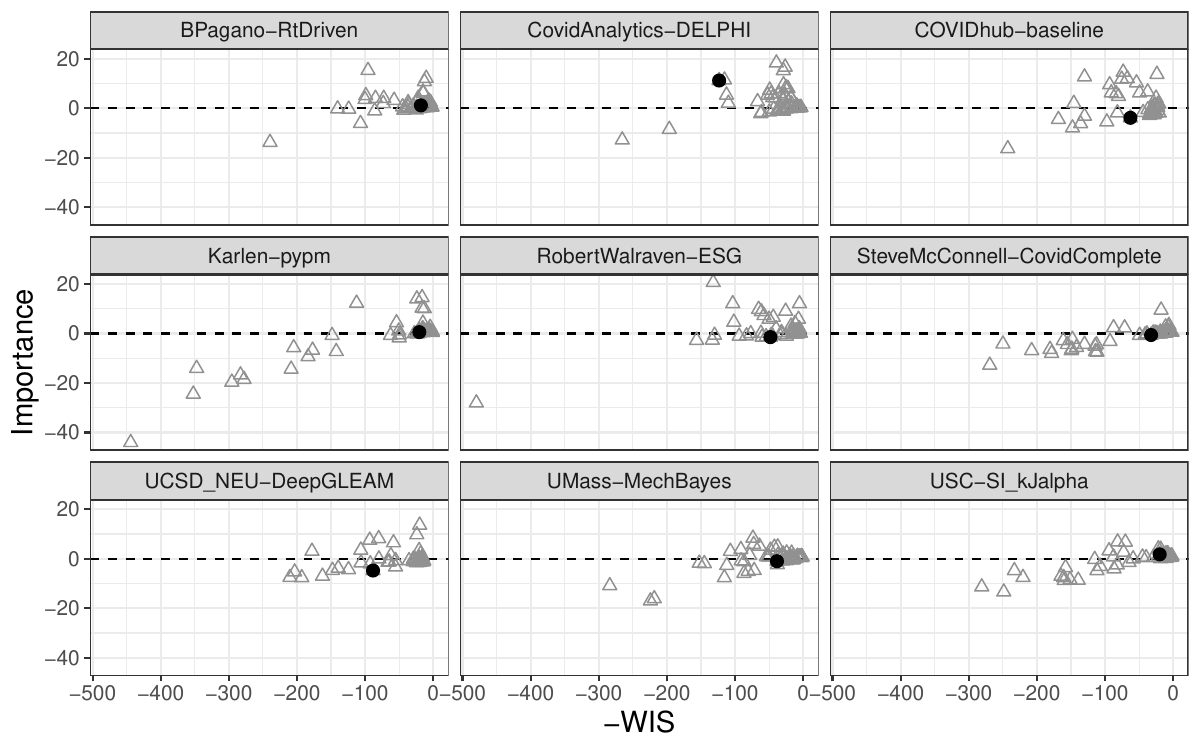}
    \caption{Model importance versus negative WIS by model for all weeks in 2021. Each triangle represents a pair of negative WIS ($x-$axis, larger values indicate more accurate forecasts) and importance metric ($y-$axis, larger values indicate more important forecasts) for a week in 2021. Solid black circles represent negative WIS and importance metric pairs evaluated for the one week ending December 25, 2021 (see more details in \hyperref[fig:20211225]{Figure \ref{fig:20211225}}). The horizontal dashed lines indicate the value of zero. The importance of an individual model as an ensemble member tends to be positively correlated with the value of negative WIS; that is, the importance metric has a positive correlation with the model's prediction accuracy.}
    \label{fig:scatterplot-wis_imp2021}
\end{figure}

\begin{figure}[b!]
    \vspace{-10ex}
    \centering
    \begin{subfigure}[c]{0.05\textwidth}
        \vspace{-23ex}
        \caption{}\label{fig:wis_imp20211225}
    \end{subfigure}%
    \begin{minipage}[c]{0.9\textwidth}
        \adjustbox{margin=0.3cm 0cm 0cm 0cm}{\includegraphics[scale=0.33]{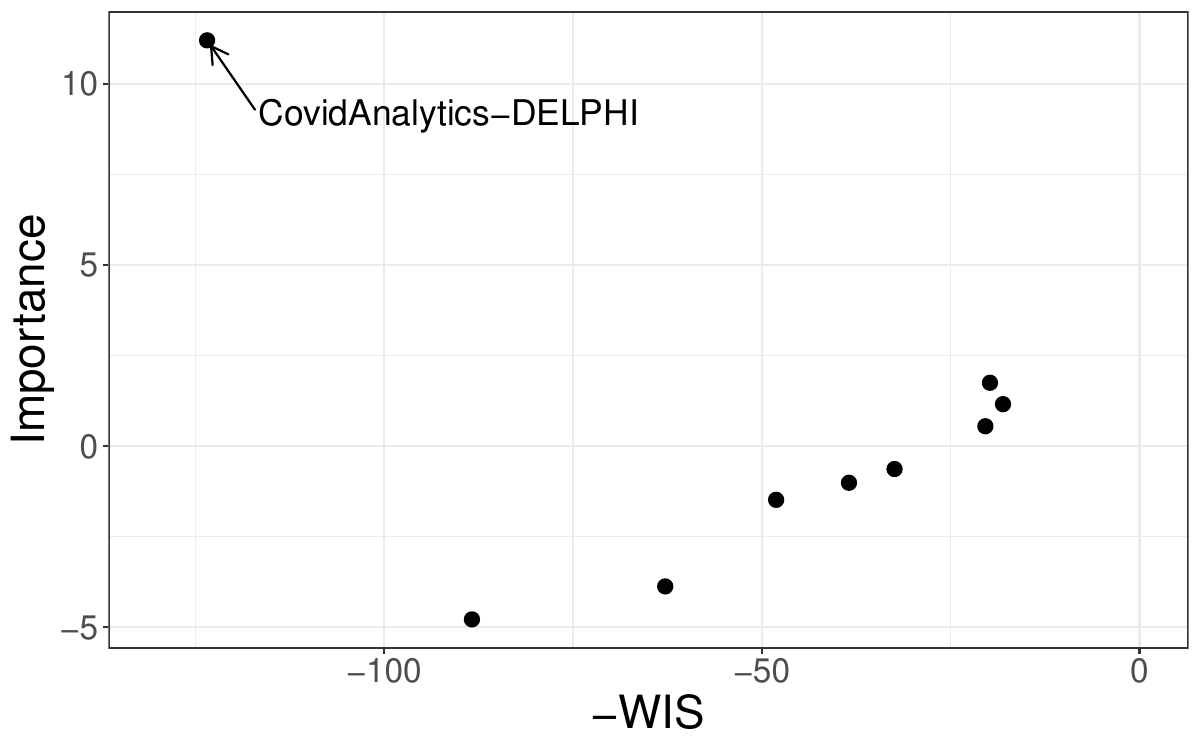}} 
    \end{minipage}
    \\[1ex]
    \begin{subfigure}[c]{0.05\textwidth}
        \vspace{-57ex}
        \caption{}\label{fig:PIs2021}
    \end{subfigure}%
    \begin{minipage}[c]{0.9\textwidth}
        \includegraphics[scale=0.55]{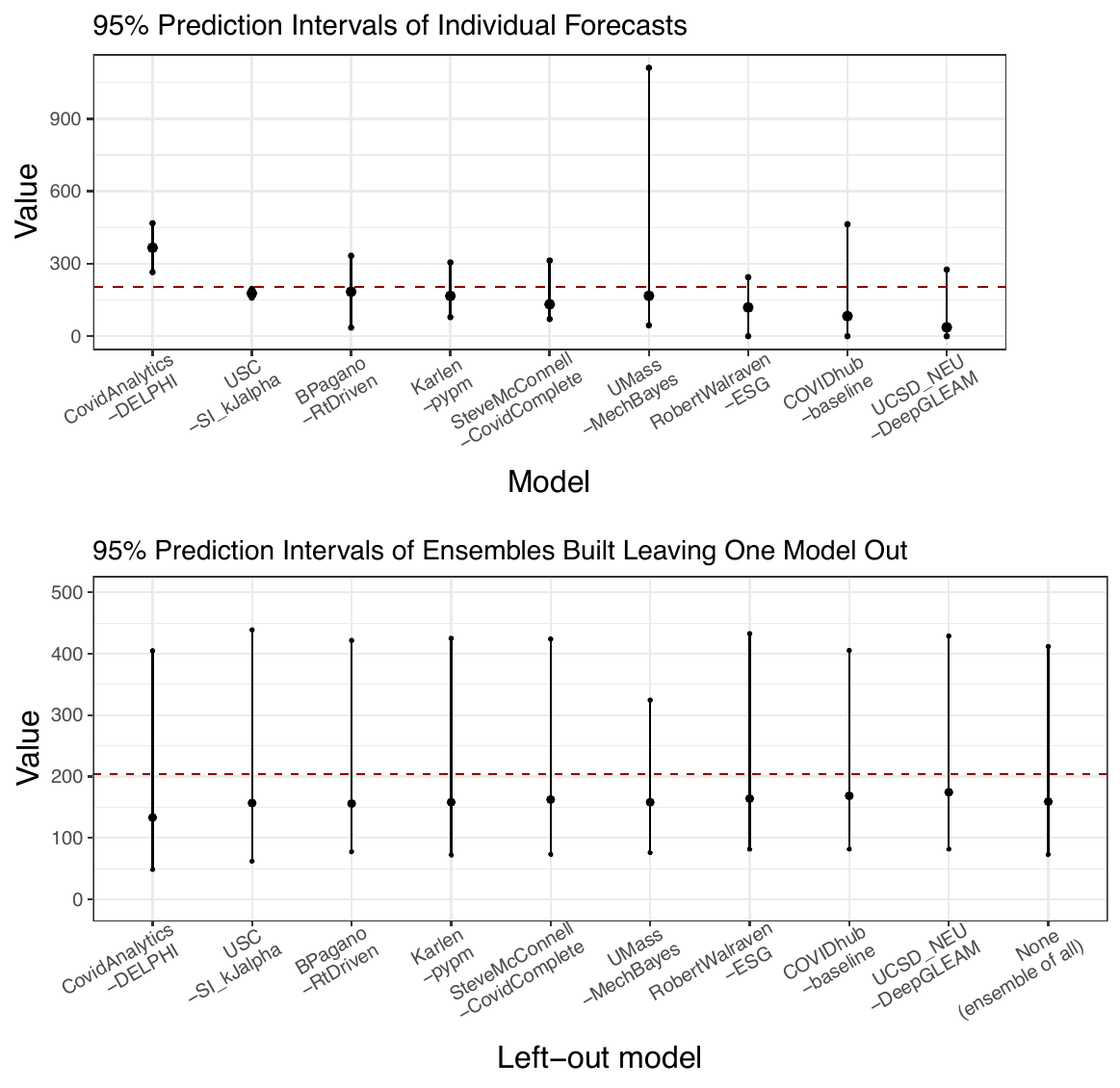}
    \end{minipage}
    \caption{(\subref{fig:wis_imp20211225}) Model importance of each model versus negative WIS in Massachusetts on target end date 2021-12-25. CovidAnalytics-DELPHI is the most important and also the least accurate by $-\text{WIS}.$ (\subref{fig:PIs2021}) Predictive medians and 95\% Prediction intervals (PIs) of individual forecasts (top) and ensemble forecasts built leaving one model out (bottom) on target end date 2021-12-25. For example, the lines on the far left indicate PI for the CovidAnalytics-DELPHI model on the top panel and PI for the ensemble created without the CovidAnalytics-DELPHI model on the bottom panel. None(ensemble of all) represents an ensemble model built on all nine individual models. In each PI, the end points indicate 0.025 and 0.975 quantiles and the mid-point represents the 0.5 quantile (predictive median). The horizontal dashed lines represent the eventual observation. The ensemble without CovidAnalytics-DELPHI is the only ensemble model with a point estimate below 150. The models on the $x$-axis are listed in order of model importance.}
    \label{fig:20211225}
\end{figure}

Our first analysis is a small case study designed to investigate the relationship between model importance calculated with the leave one model out (LOMO) algorithm and model accuracy measured by the negative WIS. Forecasts analyzed were a subset of all forecasts from the Forecast Hub, including only 4-week ahead forecasts of new deaths in Massachusetts, made for every week in 2021. The only models included were those that had made real-time forecasts for every week in 2021 to avoid handling complications that arise from missing forecasts. We also excluded models that were ensembles of other models in our pool. This led to a set of 9 individual models. In building ensemble models, an equally weighted mean at each quantile level was employed.

In Massachusetts in 2021, the importance metrics of component models were correlated with model accuracy as measured by $-\text{WIS}.$ 
Specifically, the more accurate a model's predictions were on average (as the value of negative WIS increases, it indicates higher accuracy), the higher the importance that model had (larger values indicate more important forecasts)(\hyperref[fig:scatterplot-wis_imp2021]{Figure \ref{fig:scatterplot-wis_imp2021}}).
However, there are still certain models in given weeks with high importance metrics despite having low accuracy (i.e., low value of negative WIS), which implies that there are some other factors that determine the importance of a component model.
An example is the forecasts with a target end date of December 25, 2021, where the CovidAnalytics-Delphi model was the most important contributor to the ensemble, but as measured by negative WIS it was also the least accurate for that forecast task (\hyperref[fig:wis_imp20211225]{Figure \ref{fig:wis_imp20211225}}).
This is because while this model had a large positive bias, it was the only model that was biased in that direction on this particular week (\hyperref[fig:PIs2021]{Figure \ref{fig:PIs2021}}). That bias made this model an important counter-weight to all of the other models, and adding this model to an ensemble moves the predictive median towards the observed data.
This illustrates that component forecasts that are not accurate relative to other forecasts, but that offer a unique and different perspective from other models can still play an important role in an ensemble.

\subsection{Importance metrics measured by different algorithms}\label{subsec:extensive-application}

For this application, out of a total of 56 non-ensemble individual models that submitted forecasts of COVID-19 deaths to the Forecast Hub, we chose 10 models that submitted over 90\% of the possible individual predictions for deaths across 50 states in the U.S. and 1 through 4 horizons for 109 weeks from November 2020 to November 2022 (\hyperref[tab:score-compare-NAworst]{Table \ref{tab:score-compare-NAworst}}).

As mentioned earlier, we took an equally weighted mean of the models' quantile forecasts in the ensemble construction (see \hyperref[subsec:ensemble-method]{Ensemble Methods}). 
If a model did not submit forecasts, the model's score was stored as `NA'. When compiling the scores, those `NA' values were processed in three different ways: they were excluded from the analysis, substituted with the worst specific to the combination of forecast date, location, and horizon, or substituted with the average score for the same combination. Here, we present the results obtained by adopting the most conservative approach, wherein `NA' values were replaced with the worst scores. The results from other approaches show similar patterns to the observations discussed below. The details can be found in the supplement (see Supplemental Section 4). 

\begin{table}[t!]
    \begin{center}
    \resizebox{\textwidth}{!}{
    \renewcommand*{\arraystretch}{1.4}
	\begin{tabular}{ l rrr c } 
	\hline
	Model & $-\text{WIS}$  & $\Phi^{\text{lasomo}}$ &  $\Phi^{\text{lomo}}$ & Number of predictions (\%)\\ 
	\hline 
	BPagano-RtDriven &\bftab -40.2 & 2.81  & 0.71 & 21800 (100)\\ 
	Karlen-pypm        & -41.2  &\bftab 3.11  &\bftab 0.92 & 20400 (93.6)\\ 
	GT-DeepCOVID       & -42.8  & 1.87 & 0.17 & 20724 (95.1)\\
	MOBS-GLEAM\_COVID  & -45.8  & 1.06  & -0.21 & 20596 (94.5)\\
	CU-select          & -47.3  & 1.64 & 0.24 & 21000 (96.3)\\
	RobertWalraven-ESG & -49.8  & 0.94 & -0.09 & 19992 (91.7)\\
	USC-SI\_kJalpha    & -51.7  & 1.23 & 0.21 & 20900 (95.9)\\
	COVIDhub-baseline  & -52.1  & 0.10  & -0.62 & 21800 (100)\\ 
	UCSD\_NEU-DeepGLEAM& -52.6  & -0.13 & -0.70 & 20596 (94.5)\\
	PSI-DRAFT          & -71.7  & -1.94 & -1.00 & 19988 (91.7)\\
	\hline
	\end{tabular} }
     \end{center} \vspace{-0.5cm}
     \caption{Summary of negative WIS and importance metrics ($\Phi$), sorted by $-\text{WIS}$. The number of predictions represents the total forecasts made by each model, with the percentage of the total number of predictions shown in parentheses, for the 50 US states across 1-4 week horizons from November 2020 to November 2022 (109 weeks). All scores were averaged across all forecast dates, locations, and horizons. In the importance metric notation ($\Phi$), the superscript indicates the algorithm method; $\Phi^{\text{lomo}}$ represents the average importance metric based on leave one model out algorithm and $\Phi^{\text{lasomo}}$ represents the average importance metric based on leave all subsets of models out algorithm. The best value in each column is highlighted in bold.}
      \label{tab:score-compare-NAworst}
\end{table}
      
\begin{figure}[t!]
    \centering
    \includegraphics[width=\textwidth]{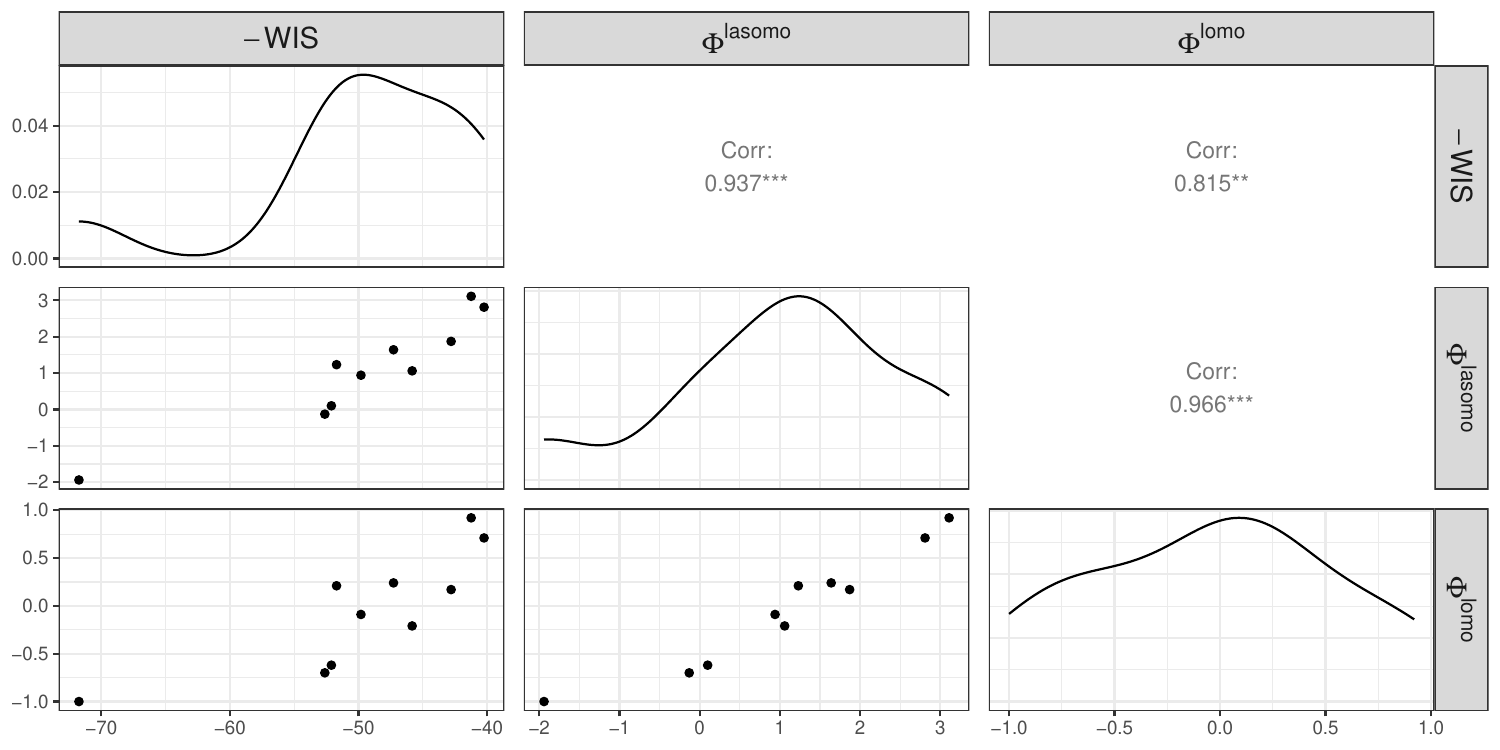}
    \caption{Relationship between summary metrics computed across the entire evaluation period. In the importance metric notation ($\Phi$), the superscript indicates the algorithm method; $\Phi^{\text{lomo}}$ represents the average importance metric based on leave one model out algorithm and $\Phi^{\text{lasomo}}$ represents the average importance metric based on leave all subsets of models out algorithm. One black dot corresponds to one model, with the position representing the average scores across the entire evaluation period for the metrics corresponding to the row and column of the plot matrix.}
    \label{fig:corr-NAworst} 
\end{figure}

Overall, the importance metrics measured through the two computational algorithms were highly correlated in the positive direction with the negative WIS (\hyperref[fig:corr-NAworst]{Figure \ref{fig:corr-NAworst}}).
That is, on average, the more accurate a model was by $-\text{WIS}$, the more important a role it played in contributing to the accuracy of an ensemble.

\begin{figure}[t!]
    \centering
    \vspace{0.45 cm}
    \begin{subfigure}{0.5\linewidth}
        \adjustbox{margin=-0.15cm 0cm 0cm 0cm}{\includegraphics[width=\linewidth]{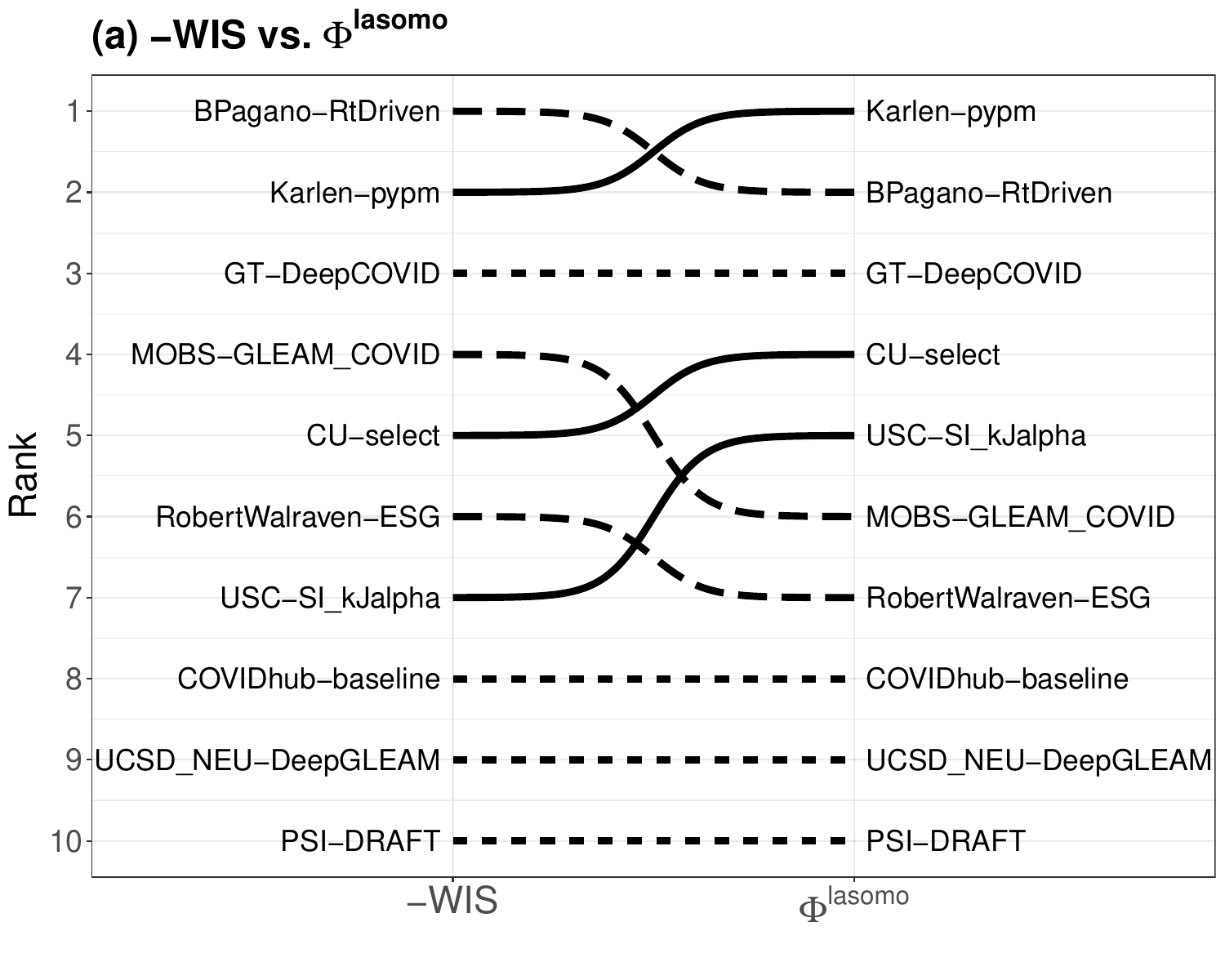}}
    \end{subfigure}\hspace{-0.25cm}
    \begin{subfigure}{0.5\linewidth}
        \adjustbox{margin=-0.15cm 0cm 0cm 0cm}{\includegraphics[width=\linewidth]{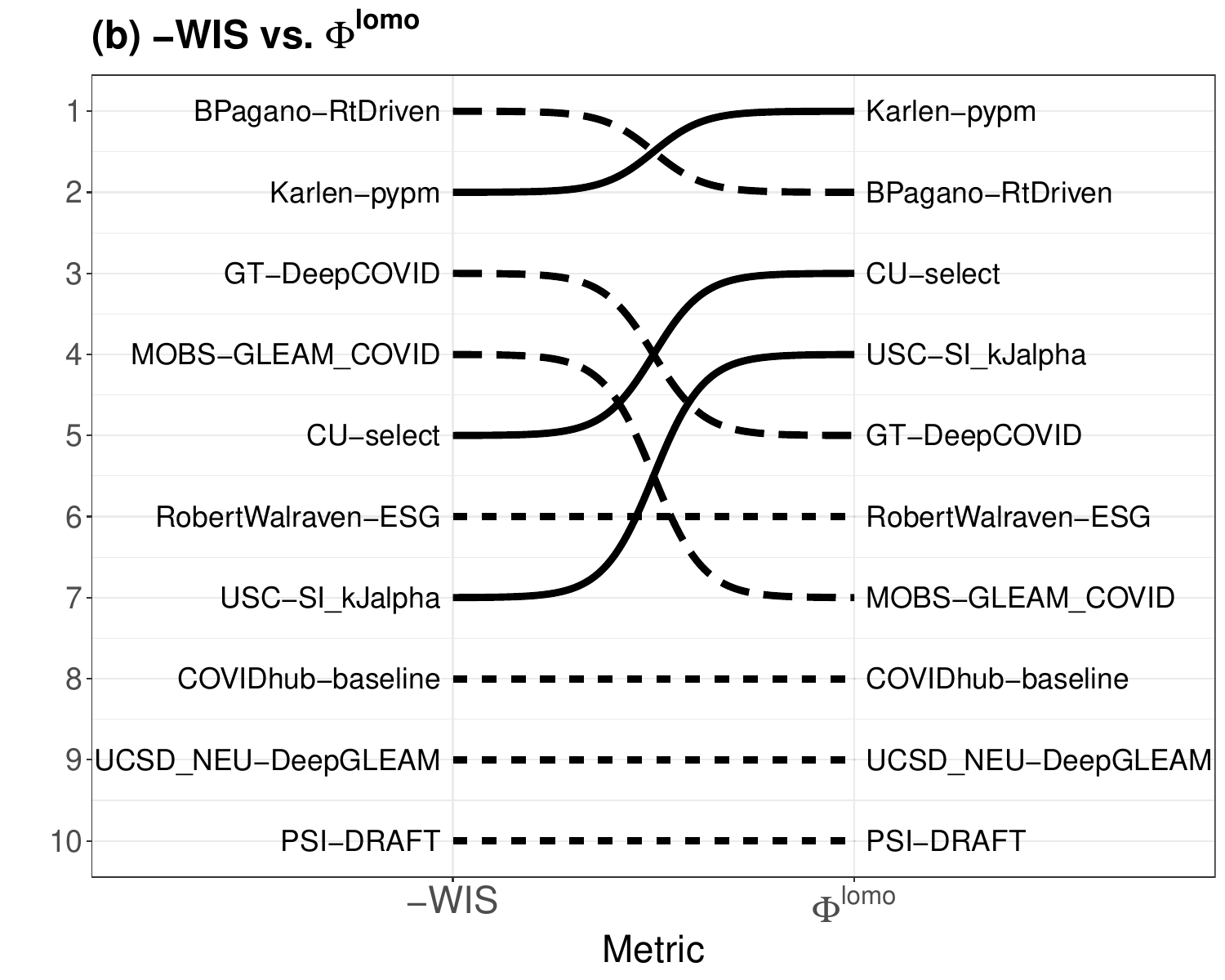}}
    \end{subfigure}\hspace{-0.25cm}
    \caption{Comparison of model ranks as measured by the negative WIS against different importance metrics: (a) $-\text{WIS}$ vs. $\Phi^{\text{lasomo}}$ and (b) $-\text{WIS}$ vs. $\Phi^{\text{lomo}}$.
    Solid lines indicate cases where the importance metric rank is higher than the negative WIS rank, dashed lines indicate lower ranks, and dotted lines represent equal ranks.}
    \label{fig:rank-changes-NAworst}
\end{figure}

In certain instances, the rankings of models by $-\text{WIS}$ or importance metrics were not the same. 
For example, the Karlen-pypm and BPagano-RtDriven models were the top two models by $-\text{WIS}$ or any of the importance metrics. Although BPagano-RtDriven showed higher accuracy by $-\text{WIS}$, Karlen-pypm showed greater importance on average, despite being substantially penalized for its missing values by assigning the worst score per the corresponding task for each metric, while BPagano-RtDriven was not penalized. This suggests that the Karlen-pypm model added more value than the BPagano-RtDriven model in its ability to meaningfully contribute to ensemble predictions.
We also observe that USC-SI\_kJalpha, which had a worse negative WIS, showed greater importance than MOBS-GLEAM\_COVID  (\hyperref[tab:score-compare-NAworst]{Table \ref{tab:score-compare-NAworst}}, \hyperref[fig:rank-changes-NAworst]{Figure \ref{fig:rank-changes-NAworst}}), where the penalties incurred by both models were comparable. 
This implies that even models with low accuracy as measured by $-\text{WIS}$ can provide a unique perspective that differentiates them from other models as standalone predictive models, and thereby further contribute to improving the average ensemble.

Factors that are not captured by $-\text{WIS}$ but influence the importance metric can be explained by the similarity between models. In \hyperref[eq:mean-phii]{Eq.(\ref{eq:mean-phii})}, the importance metric is decomposed into individual forecast skills and the similarities of forecast errors from different component models for point forecasts. This concept can also be applied to probabilistic quantile-based forecasts, as demonstrated in \hyperref[subsubsec:simu-settingA]{Section \ref{subsubsec:simu-settingA}}. This similarity is understood in terms of how often the prediction errors fall ``on the same side" of the observation and how much a particular model corrects errors from other models. 

In general, the importance metrics for different computational algorithms (LASOMO/LOMO) are highly correlated with each other (\hyperref[fig:corr-NAworst]{Figure \ref{fig:corr-NAworst}}). The relative ordering of models is not particularly sensitive to this choice. However, the importance metric calculated in LOMO, denoted by $\Phi^{\text{lomo}}$, is consistently lower than the importance metric calculated in LASOMO, denoted by $\Phi^{\text{lasomo}}$, for each model. Notably, many models that had positive scores in the LASOMO approach exhibit negative scores in LOMO approach (\hyperref[tab:score-compare-NAworst]{Table \ref{tab:score-compare-NAworst}}). This can be interpreted that it is harder for a model to add value when all other models are already in the mix. It is because  $\Phi^{\text{lomo}}$ represents the model's marginal contribution to the subset that includes all the other models, and it is considered only a part of $\Phi^{\text{lasomo}}$.
\\

As a complementary analysis, we also explored the variability of LOMO metrics across different subset sizes, which illustrates their influence on LASOMO metrics (see Supplement, Section 5). We found that subsets with a small number of models sometimes exhibit high variance in the LOMO metrics, which can lead to potential instability in the LASOMO metrics. On the other hand, the LOMO model importance scores for a particular model are generally stable when one or two other models are removed from the pool of models considered.

\section{Discussion}\label{sec:discussion}

We have developed the importance metric using the Shapley value concept. While earlier studies applied this concept to measure the contribution of individual predictors of a predictive model (\citet{Giudici2021, borup2024anatomy}) or ensemble component models (\citet{Pompigna2018, Adiga2023}) in terms of point prediction accuracy, we have explored the concept in the context of probabilistic ensemble forecasts of epidemics. We also provided a detailed understanding of the accuracy and similarity decomposition in the model importance metric, 
revealing the conditions under which a component model is rewarded in the presence of other models. This highlights a key distinction from the ambiguity decomposition proposed by \citet{brown2005}.

The Weighted Interval Score (WIS), a proper scoring rule designed for quantile forecasts, was utilized by the US COVID-19 Forecast Hub and several other collaborative forecasting challenges (\citet{Mathis2024, Sherratt2023, Howerton2023}). While WIS is effective in measuring each model's performance independently, it does not offer a complete picture of a model's contribution in an ensemble setting. The importance metric approaches that we introduce in this work provide insights that cannot be captured by WIS, as it is contingent upon predictions from other models. No matter how accurate a prediction is, if its prediction errors are highly similar to those of other models, its impact on the ensemble may not be as great as that of a model that has lower accuracy but offsets the errors of other models. This aspect of importance metrics is especially relevant for hub organizers like the US CDC, as they collect forecasts from a variety of models and combine them to generate ensemble forecasts for communication with the public and decision-makers (\citet{Fox2024}). 

We proposed two algorithms for assessing model importance: leave-one-model-out (LOMO) and leave-all-subsets-of-models-out (LASOMO). 
In the LASOMO algorithm, we used permutation-based weights to account for how a model's contribution can vary depending on the ensemble size, which distinguishes our work from that of \citet{Adiga2023}, who use an equal-weighting scheme. 
Notably, LOMO is a special case of LASOMO in that it considers only a single subset, which includes all models except the target component model. This makes LOMO simple and easy to implement and significantly more efficient than LASOMO, especially when dealing with many component models. However, when the number of component models is relatively small (e.g., fewer than 10), LASOMO becomes computationally feasible, and thus it may be preferred, as it provides a more comprehensive evaluation by considering all possible combinations of component models.  

This study has several limitations. While we utilized the widely adopted mean ensemble method in the application, this method is often vulnerable to individual forecasters with outlying or poorly calibrated predictions, which can increase forecast uncertainty or decrease overall reliability in the ensemble. Additionally, our use of Shapley values, while providing insights, was constrained by underlying assumptions, such as assigning a zero value to the empty set in the characteristic function, that were not fully met in our setting. Thus, the obtained values may not precisely represent Shapley values but rather provide a informal approximation of them. The computational cost of implementing the LASOMO algorithm is also a challenge. As the number of models increases, the computational time grows exponentially because a total of $2^n$ subsets must be considered for $n$ models in the Shapley value calculation. 
Furthermore, it is almost impossible to have all models consistently submit their predictions over a given period when there are many participating models, so it is inevitable to have missing data for unsubmitted predictions. Consequently, the Shapley value can be unstable or misleading, as it is highly sensitive to such missingness. As the choice of how to handle these missing values during scoring can lead to variations in the resulting importance metrics and rankings of the component models, careful consideration is required when choosing the handling method. Further exploration of this issue is needed for comprehensive guidelines.

We also envision several directions for future research. A naïve forecast could be incorporated as a baseline model, serving as a replacement for the forecast associated with the empty set of component models, which we excluded in this study.
Another potential direction is to explore the application of model importance measures in the context of ensemble forecasts that assign weights to individual component models. In this case, more deliberate strategies should be explored to account for the different levels of consistency and reliability across models in the weighting scheme (\citet{Ray2023}). 
Moreover, the components of the LASOMO metric computed using subsets of a few models sometimes exhibit high variance.  While this is less of an issue for LOMO metrics when a reasonable number of models are available, several approaches could be explored  to reduce the impact of high-variance metrics and outliers. We could refine the LASOMO framework to develop variance-adjusted versions of the metric. It could also be valuable to explore ways to extend the Rank Graduation Accuracy (RGA) metrics used in the point predictions (\citet{giudici2025rga}) in the context of a probabilistic forecasting setting. Such a probabilistic implementation of RGA could improve robustness, as it would be less affected by outliers.

\section{Conclusions}\label{sec:conclusions}
Despite the rising popularity of ensemble models, there is currently a lack of comprehensive evaluation metrics to assess the individual contributions of each model. Traditional standard practice involves setting up a leaderboard to independently evaluate the accuracy and effectiveness of individual prediction models based on appropriate scoring rules. Our proposed importance metric addresses this gap by offering a novel and distinctive metric for assessing the role of each model within the ensemble, adding a unique dimension to the assessment of forecasting models.

This paper presents a decomposition of the model importance metric, which mathematically demonstrates how an individual model’s accuracy and its interactions with other component models influence the measure. This theoretical framework is supported by simulation studies. In a case study, its application is illustrated in a real-world setting. These analyses provide both formal and intuitive explanations of the realized values of the model importance metrics, and highlight ways in which the model importance metric can be used to understand how individual models have improved or degraded predictive accuracy. An extensive application further highlights the relationship between a widely used accuracy metric and our model importance metric. 

The implication of this work is that our proposed importance metric provides novel insights, offering new information beyond traditional accuracy metrics. Our method provides a solid theoretical basis and clear criteria for quantifying a component model's contribution to ensemble performance. Moreover, leveraging the importance metric has the potential to incentivize original modeling approaches, thereby fostering a diverse landscape of perspectives among modelers, ultimately enriching the forecasting ecosystem.

\section{Declaration of Competing Interest}
The authors declare that they have no known competing financial interests or personal relationships that could have appeared to influence the work reported in this paper. 

\section{Acknowledgments}\label{sec:acknowledgments}

We acknowledge Daniel Sheldon for helping seed the original idea for this work. This study started by taking the kernel of his ideas about having models ranked based on their unique contribution to making the ensemble better.
Conversations with Mark Wilson were also helpful in the nascent stages of thinking about and understanding Shapley values.
This work has been supported by the National Institutes of General Medical Sciences (R35GM119582) and the US CDC (1U01IP001122). The content is solely the responsibility of the authors and does not necessarily represent the official views of CDC, NIGMS or the National Institutes of Health.

\appendix
\section{Supplementary data}
\label{appendix}
The following is the Supplementary material related to this article.

\bibliography{references}
\end{document}


\maketitle
\section{An interpretation of the Shapley value} \label{supp_sec:Shapley_value_interpretation}
An interpretation of the Shapley value (\citet{Shapley1953}) is given as follows. For a given player $i$ among $n$ players, say $i=1$, 
we can construct subsets of players that do not include player $i$ and categorize them according to the subset size indicated by the subscript, as follows:
\begin{align}\label{list:group_subset}
	G_0 &=\big[\emptyset \big] \nonumber \\
	G_1 &=\big[ \{2\}, \{3\}, \{4\}, \cdots, \{n-1\}, \{n\}\big] \nonumber \\
	G_2 &=\big[ \{2,3\}, \{2,4\}, \{2,5\}, \cdots, \{n-2, n-1\}, \{n-2, n\}, \{n-1, n\}\big]  \\
	& \vdots \nonumber \\
	G_{n-1} &= \big[ \{2,3,4,5,\cdots, n-1, n\} \big] \nonumber 
\end{align}
Then, the cardinality of $G_s, \,(s\in \{0,1,2, \cdots, n-1\}),$ is $$|G_s|= \binom{n-1}{s} =\frac{(n-1)!}{s!(n-1-s)!},$$ 
and the Shapley value is the expected contribution of player $i$ to a particular coalition $\tilde S \in G_s$, because the probability of getting the coalition $\tilde S$ is computed as 
\begin{align} \label{eq:prob_getting_coalition}
	&Pr(G=G_s) Pr(S = \tilde S |G_s) \nonumber\\ 
	=& \frac{1}{n}\cdot\frac{1}{|G_s|} =\frac{1}{n}\cdot\frac{s!(n-s-1)!}{(n-1)!} = \frac{s!(n-s-1)!}{n!},
\end{align}
which is the multiplier in Eq. (5) of the main text. 

If one excludes the case of the empty set of forecasts,  the subset $G_0$ in the \hyperref[list:group_subset]{List(\ref{list:group_subset})} is removed from consideration, and consequently, $Pr(G=G_s)$ in \hyperref[eq:prob_getting_coalition]{Eq.(\ref{eq:prob_getting_coalition})} is replaced with $1/(n-1)$.

\section{Importance metric calculation}\label{supp_sec:derivation}
The derivation steps between Eq. (9) and Eq. (10) of the main text are given as follows.
\begin{align}\label{supp-eq:phii}
    \phi_i &= -\left(y-\frac{1}{n}\sum_{j=1}^{n} \hat{y}_j\right)^2 + \left(y-\frac{1}{(n-1)}\sum_{j\ne i} \hat{y}_j\right)^2 \\
    &=-\left(\frac{1}{n}\sum_{j=1}^{n} (y-\hat{y}_j)\right)^2+ \left(\frac{1}{(n-1)}\sum_{j\ne i} (y-\hat{y}_j)\right)^2\nonumber\\
    &=-\frac{1}{n^2}\left(\sum_{j=1}^{n} e_j\right)^2+\frac{1}{(n-1)^2} \left(\sum_{j\ne i} e_j\right)^2 \nonumber\\
    & = -\frac{1}{n^2}\left(e_i + \sum_{j\ne i} e_j\right)^2+\frac{1}{(n-1)^2} \left(\sum_{j\ne i} e_j\right)^2 \nonumber\\
    & = -\frac{1}{n^2}\left(e_i^2 + 2e_i\sum_{j\ne i} e_j + \bigg(\sum_{j\ne i} e_j\bigg)^2\right)+\frac{1}{(n-1)^2} \left(\sum_{j\ne i} e_j\right)^2 \nonumber\\
    & = -\frac{1}{n^2} e_i^2 -\frac{2}{n^2}e_i\sum_{j\ne i} e_j - \left(\frac{1}{n^2}-\frac{1}{(n-1)^2}\right) \bigg(\sum_{j\ne i} e_j\bigg)^2  \nonumber\\
    & = -\frac{1}{n^2} e_i^2 -\frac{2}{n^2}\sum_{j\ne i} e_ie_j + \frac{2n-1}{[n(n-1)]^2} \left(\sum_{j\ne i} e_j^2 + 2\sum_{\substack{j\ne i\\ j<k}} e_je_k\right). \nonumber
\end{align}

\subsection{Connection to the ambiguity decomposition}\label{supp_sec:connetion-to-ambiguity-decomp}
Our model importance metric decomposition is closely related to the ambiguity decomposition of \citet{brown2005} in the context of point forecasts. Using their notation, the ambiguity decomposition is expressed as
\begin{equation}\label{supp-eq:decomp-full}
    (f_{\text{ens}}-d)^2 = \sum_j w_j(f_j-d)^2-\sum_j w_j(f_j-f_{\text{ens}})^2,
\end{equation}
where $f_{\text{ens}}$ is a weighted ensemble, $f_j$ is an ensemble component with ensemble weight $w_j$, and $d$ is the actual outcome. The first term represents the weighted average error of the component models and the second term is called the ambiguity term. 

Let $f_{\text{ens}}^{-i}$ denote an ensemble leaving out model $i$. The corresponding decomposition is 
\begin{equation}\label{supp-eq:decomp-no-i}
    (f_{\text{ens}}^{-i}-d)^2 = \sum_{j\ne i} w_j(f_j-d)^2-\sum_{j\ne i} w_j(f_j-f_{\text{ens}}^{-i})^2.
\end{equation}
Subtracting Eq. (\ref{supp-eq:decomp-no-i}) from Eq. (\ref{supp-eq:decomp-full}) yields $\phi_i$ on the left-hand side, by the definition as in Eq. (\ref{supp-eq:phii}). The right-hand side simplifies to
\begin{align*}
    -w_i e_i^2 + \sum_{j} w_j(f_j-f_{\text{ens}}^{-i})^2 -\sum_{j\ne i} w_j(f_j-f_{\text{ens}}^{-i})^2,
\end{align*}
where $e_i=f_i-d$.
Hence, the difference in the ambiguity terms and our importance metric differ only by the model $i$'s squared error, scaled by its ensemble weight.

\section{Standard deviation of simple mean ensembles}\label{supp_sec:std-ensembles}

Consider a probabilistic forecast represented as a set of $r$ predictive quantiles $q_k$ at levels $\tau_k,$ $k=1,2, ..., r,$ which is normally distributed with mean $0$ and variance $s^2$
$$ F = N(0, s^2).$$
Then, predictive quantiles can be expressed in terms of the standard deviation $s$ and quantiles of the unit normal distribution. For example, at $\tau_k=0.975$ for some $k,$
\begin{equation}\label{eq:qk_std}
    q_k=1.96\times s \equiv \beta_k\times s,
\end{equation}
where $\beta_k=1.96$ is the 0.975 quantile of the unit normal distribution. 

We use this expression (\ref{eq:qk_std}) to find the variance of ensembles. 
If there are $N\, (N\ge2)$ normally distributed forecasts with the mean $0$ and variance $s_i^2$:
$$F_{i} = N(0, s_i^2), \quad(i=1,...,N),$$
then for a single value of $k$, the corresponding $k$th quantile of the forecast $i$, $q_{i,k}$, can be written as
$$q_{i,k}=\beta_{k}\times s_i, \quad(i=1,...,N),$$
and the simple mean ensemble built from all the $N$ models has the $k$th quantile, $q_{e,k}$, of the form 
$$q_{e,k}=\frac{\beta_{k}}{N} \sum_{j=1}^N s_j.$$
This implies that the standard deviation of the ensemble is  $(s_1+\cdots+s_N)/N.$ This is a special case of the general result that a quantile average ensemble of distributions from the same location-scale family also falls in that location-scale family, with location and scale parameters that are the average of the parameters of the individual distributions (\citet{thomas1980}).

\section{Alternative handling of NA values in importance metric calculation}\label{supp_sec:alternative-NA}

We present application results obtained by handling `NA' values using alternative approaches, different from the way we handled them in Section 3.4 of the main text: either excluding them from the analysis or substituting them with the average score for the corresponding combination of forecast date, location, and horizon.

\begin{figure}[H]
    \centering
    \includegraphics[width=\textwidth]{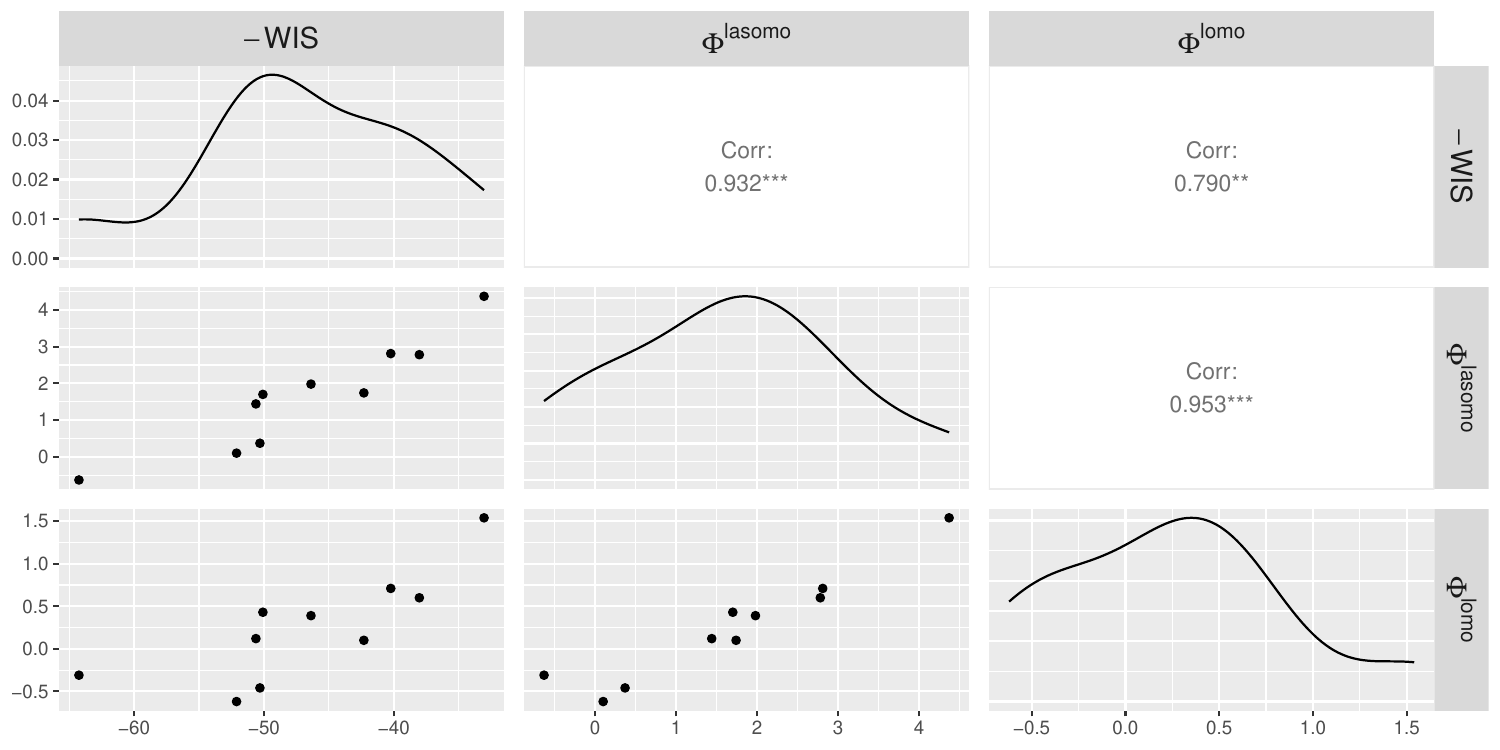}
    \caption{Relationship between summary metrics computed across the entire evaluation period. In the importance metric notation ($\Phi$), the superscript indicates the algorithm method; $\Phi^{\text{lomo}}$ represents the average importance metric based on leave one model out algorithm and $\Phi^{\text{lasomo}}$ represents the average importance metric based on leave all subsets of models out algorithm . One black dot corresponds to one model, with the position representing the average scores across the entire evaluation period for the metrics corresponding to the row and column of the plot matrix.}
    \label{fig:corr-NAdrop} 
\end{figure}

\begin{figure}[h!]
    \centering
    \includegraphics[width=\textwidth]{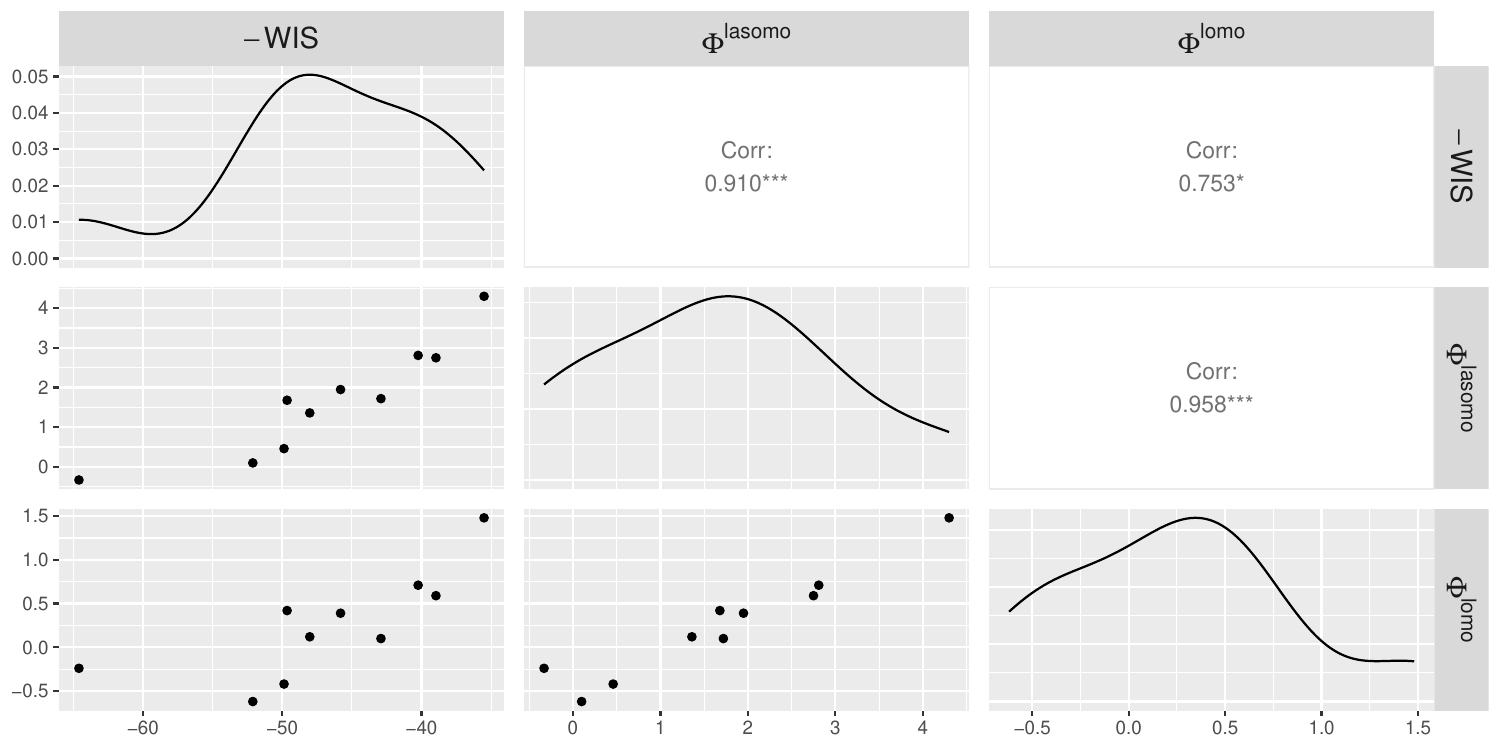}
    \caption{Relationship between summary metrics computed across the entire evaluation period. In the importance metric notation ($\Phi$), the superscript indicates the algorithm method; $\Phi^{\text{lomo}}$ represents the average importance metric based on leave one model out algorithm and $\Phi^{\text{lasomo}}$ represents the average importance metric based on leave all subsets of models out algorithm . One black dot corresponds to one model, with the position representing the average scores across the entire evaluation period for the metrics corresponding to the row and column of the plot matrix.}
    \label{fig:corr-NAavg} 
\end{figure}

In these approaches, unlike what is presented in the main text, where heavy penalties were applied, the best model by $-\text{WIS}$ consistently achieved notably high scores across all importance metrics because the absence or reduction of penalties was applied. Aside from this difference, the results from these alternative approaches align closely with the results shown in the main text. Overall, the importance metrics calculated using the two computational algorithms, demonstrated strong positive correlations with the negative WIS (\hyperref[fig:corr-NAdrop]{Figure \ref{fig:corr-NAdrop}, \ref{fig:corr-NAavg}}). In addition, the model rankings according to $-\text{WIS}$ and importance metrics were not aligned, and $\Phi^{\text{lomo}}$ consistently showed lower values than $\Phi^{\text{lasomo}}$ for each model (\hyperref[tab:score-compare-NAdrop]{Table \ref{tab:score-compare-NAdrop}, \ref{tab:score-compare-NAavg}}).

\newpage
\subsection{Dealing with NA values by dropping}

\begin{table}[h!]
    \begin{center}
    \resizebox{\textwidth}{!}{
    \renewcommand*{\arraystretch}{1.4}
	\begin{tabular}{ l r r r c } 
	\hline
	Model & $-\text{WIS}$  & $\Phi^{\text{lasomo}}$ & $\Phi^{\text{lomo}}$ & Number of predictions (\%)\\
	\hline 
	Karlen-pypm  & \bftab -33.06  &\bftab  4.37 &\bftab  1.54 & 20400 (93.6)\\ 
	GT-DeepCOVID  & -38.04 &  2.78  &  0.60 & 20724 (95.1)\\ 
	BPagano-RtDriven &  -40.24 & 2.81  &   0.71  & 21800 (100)\\
	MOBS-GLEAM\_COVID  & -42.31  & 1.74 & 0.10 & 20596 (94.5)\\
	CU-select  & -46.38  &  1.98  & 0.39  & 21000 (96.3)\\
	USC-SI\_kJalpha  & -50.03  &   1.65 & 0.41 & 20900 (95.9)\\
	UCSD\_NEU-DeepGLEAM & -50.31  & 0.37 & -0.46 & 20596 (94.5)\\
	RobertWalraven-ESG & -50.62 & 1.44 & 0.12  & 19992 (91.7)\\ 
	COVIDhub-baseline & -52.11 & 0.10  & -0.62 & 21800 (100)\\
	PSI-DRAFT& -64.24 & -0.63  & -0.31 & 19988 (91.7)\\
	\hline
	\end{tabular} }\vspace{-0.5cm}
    \end{center}
    \caption{Summary of negative WIS and importance metrics ($\Phi$), sorted by $-\text{WIS}$. The number of predictions represents the total forecasts made by each model, with the percentage of the total number of predictions shown in parentheses, for the 50 US states across 1-4 week horizons from November 2020 to November 2022 (109 weeks). All scores were averaged across all forecast dates, locations, and horizons. In the importance metric notation ($\Phi$), the superscript indicates the algorithm method; $\Phi^{\text{lomo}}$ represents the average importance metric based on leave one model out algorithm and $\Phi^{\text{lasomo}}$ represents the average importance metric based on leave all subsets of models out algorithm. The best value in each column is highlighted in bold.}
    \label{tab:score-compare-NAdrop} 
\end{table}

\begin{figure}[H]
    \centering
    \begin{subfigure}{0.5\linewidth}
        \adjustbox{margin=-0.15cm 0cm 0cm 0cm}{\includegraphics[width=\linewidth]{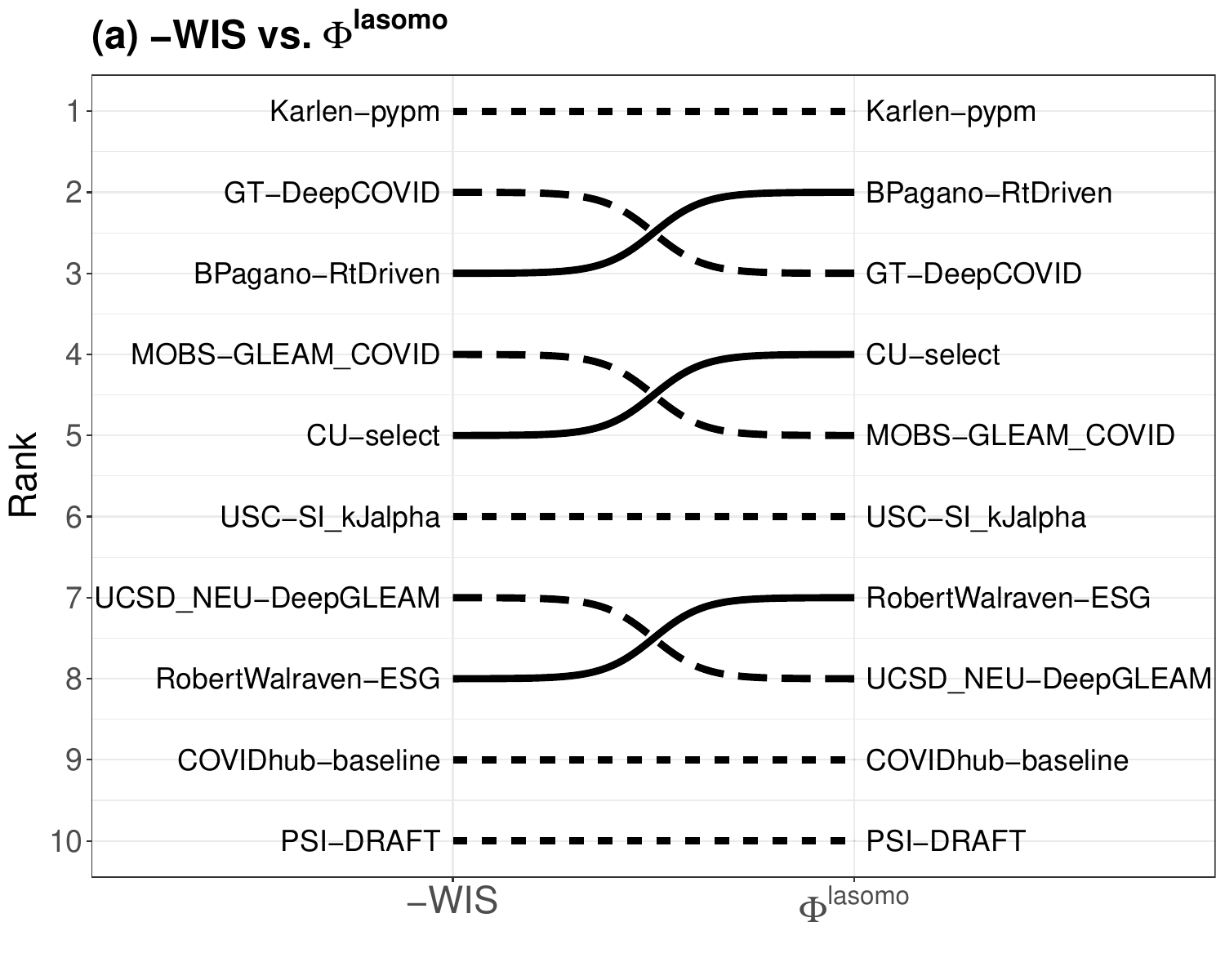}}
    \end{subfigure}\hspace{-0.25cm}
    \begin{subfigure}{0.5\linewidth}
        \adjustbox{margin=-0.15cm 0cm 0cm 0cm}{\includegraphics[width=\linewidth]{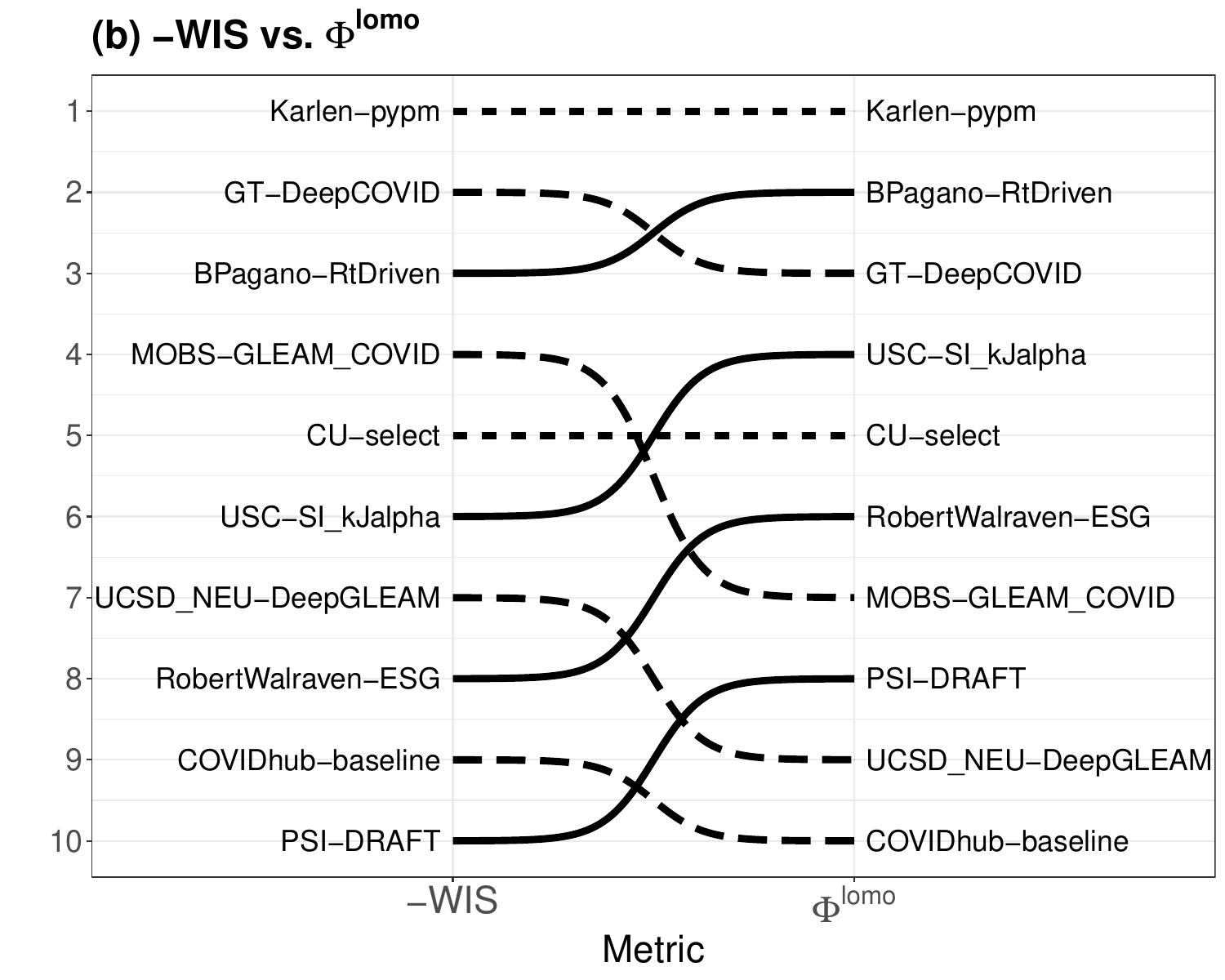}}
    \end{subfigure}\hspace{-0.25cm}
    \caption{Comparison of model ranks as measured by the negative WIS against different importance metrics: (a) $-\text{WIS}$ vs. $\Phi^{\text{lasomo}}$ and (b) $-\text{WIS}$ vs. $\Phi^{\text{lomo}}$.
    Solid lines indicate cases where the importance metric rank is higher than the negative WIS rank, dashed lines indicate lower ranks, and dotted lines represent equal ranks.}
    \label{fig:rank-changes-NAdrop}
\end{figure}

\newpage
\subsection{Dealing with NAs by mean replacement}

\begin{table}[h!]
    \begin{center}
    \resizebox{\textwidth}{!}{
    \renewcommand*{\arraystretch}{1.4}
	\begin{tabular}{ l r r r c }  
	\hline
	Model & $-\text{WIS}$  & $\Phi^{\text{lasomo}}$ & $\Phi^{\text{lomo}}$ & Number of predictions (\%)\\
	\hline 
	Karlen-pypm  & \bftab -35.51  & \bftab 4.30 & \bftab 1.48 & 20400 (93.6)\\ 
	GT-DeepCOVID  & -38.96 & 2.75 & 0.59  & 20724 (95.1)\\ 
	BPagano-RtDriven &  -40.24 & 2.81  & 0.71  & 21800 (100)\\
	MOBS-GLEAM\_COVID  & -42.91 & 1.72   &  0.10  & 20596 (94.5)\\
	CU-select  & -45.81 &  1.95  & 0.39  & 21000 (96.3)\\
	RobertWalraven-ESG &  -48.02 & 1.36  &  0.12 & 19992 (91.7)\\
	USC-SI\_kJalpha &  -49.65  & 1.68  &  0.42  & 20900 (95.9)\\
	UCSD\_NEU-DeepGLEAM &  -49.87 & 0.46  & -0.42  & 20596 (94.5)\\ 
	COVIDhub-baseline &  -52.11 &  0.10  & -0.62  & 21800 (100)\\
	PSI-DRAFT&   -64.59  &  -0.33  & -0.24 & 19988 (91.7)\\
	\hline
	\end{tabular} }
    \end{center}\vspace{-0.5cm}
    \caption{Summary of negative WIS and importance metrics ($\Phi$), sorted by $-\text{WIS}$. The number of predictions represents the total forecasts made by each model, with the percentage of the total number of predictions shown in parentheses, for the 50 US states across 1-4 week horizons from November 2020 to November 2022 (109 weeks). All scores were averaged across all forecast dates, locations, and horizons. In the importance metric notation ($\Phi$), the superscript indicates the algorithm method; $\Phi^{\text{lomo}}$ represents the average importance metric based on leave one model out algorithm and $\Phi^{\text{lasomo}}$ represents the average importance metric based on leave all subsets of models out algorithm. The best value in each column is highlighted in bold.}
    \label{tab:score-compare-NAavg} 
\end{table}

\begin{figure}[h!]
    \centering
    \begin{subfigure}{0.5\linewidth}
        \adjustbox{margin=-0.15cm 0cm 0cm 0cm}{\includegraphics[width=\linewidth]{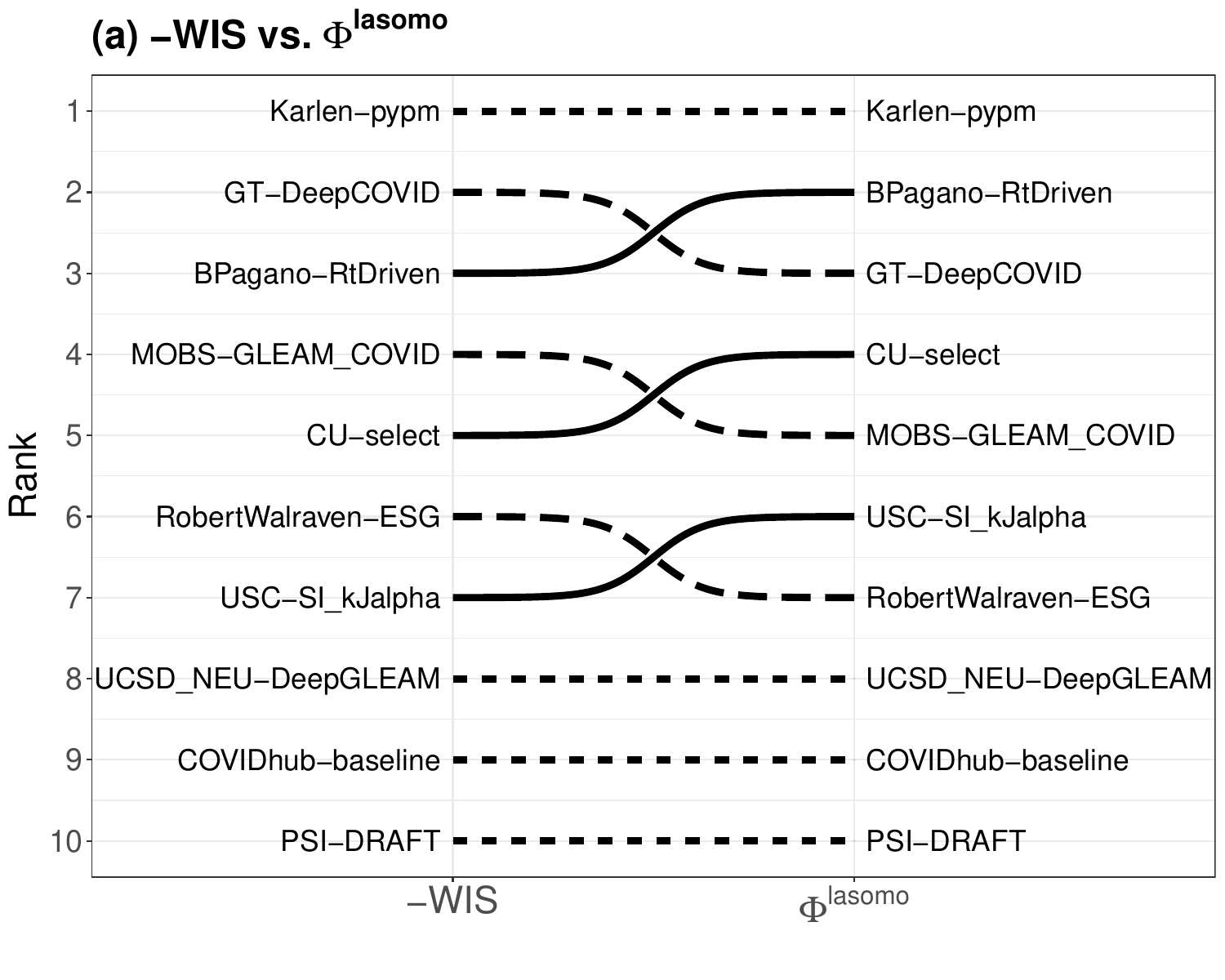}}
    \end{subfigure}\hspace{-0.25cm}
    \begin{subfigure}{0.5\linewidth}
        \adjustbox{margin=-0.15cm 0cm 0cm 0cm}{\includegraphics[width=\linewidth]{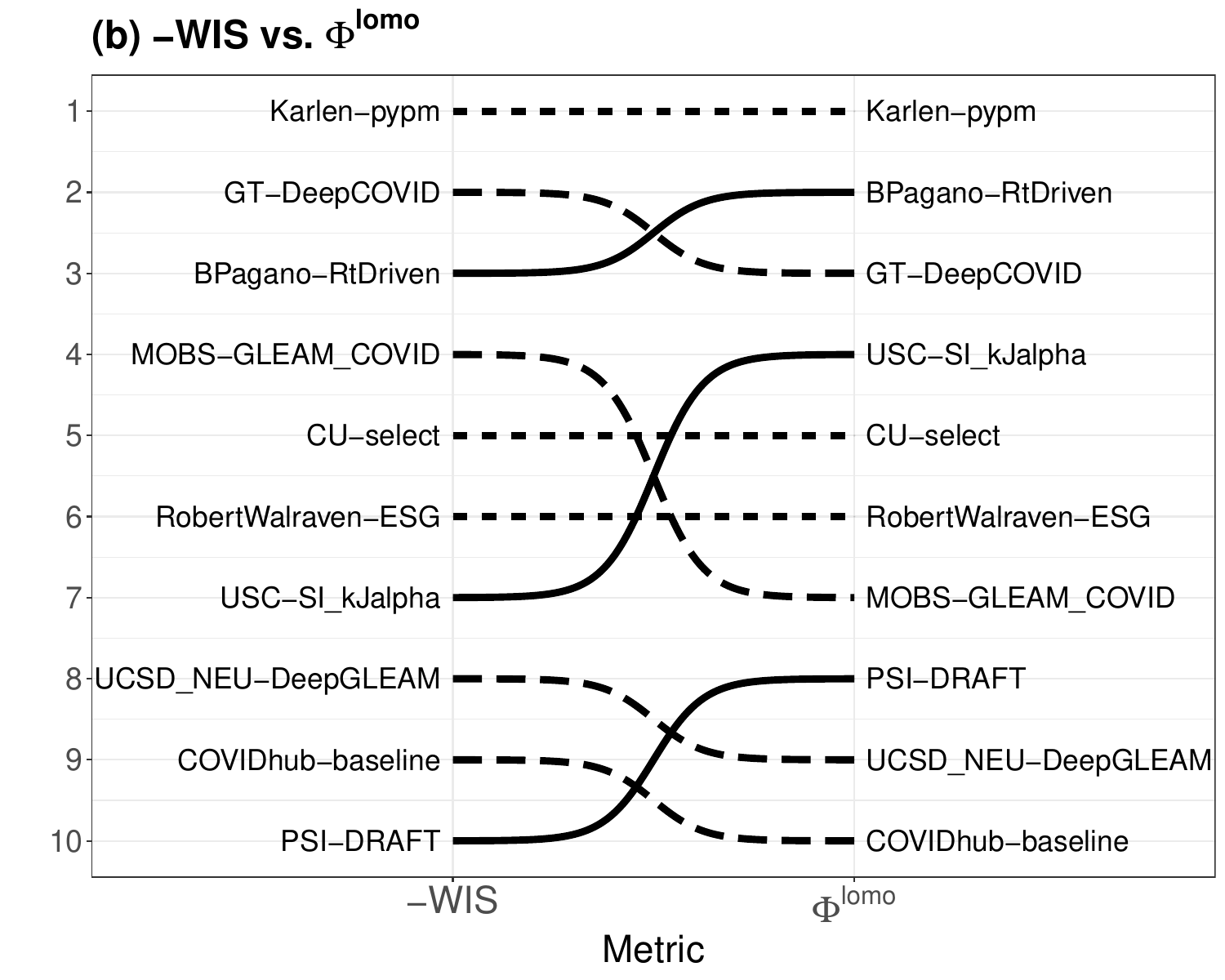}}
    \end{subfigure}\hspace{-0.25cm}
    \caption{Comparison of model ranks as measured by the negative WIS against different importance metrics: (a) $-\text{WIS}$ vs. $\Phi^{\text{lasomo}}$ and (b) $-\text{WIS}$ vs. $\Phi^{\text{lomo}}$.
    Solid lines indicate cases where the importance metric rank is higher than the negative WIS rank, dashed lines indicate lower ranks, and dotted lines represent equal ranks.}
    \label{fig:rank-changes-NAavg}
\end{figure}

\newpage
\section{Variance analysis across subset sizes}

We explored the robustness of the model importance metric on the case study of 4-week ahead incident death forecasts in Massachusetts in 2021. Specifically, we aimed to investigate how the LOMO metric varies across a range of subset sizes to better characterize the robustness of both LOMO and LASOMO metrics.

\begin{figure}[H]
    \centering
    \begin{subfigure}[b]{\textwidth}
        \captionsetup{position=above} 
        \caption{Model importance metric distribution (Target end date: 2021-07-24)}
        \includegraphics[page=1, width=\textwidth]{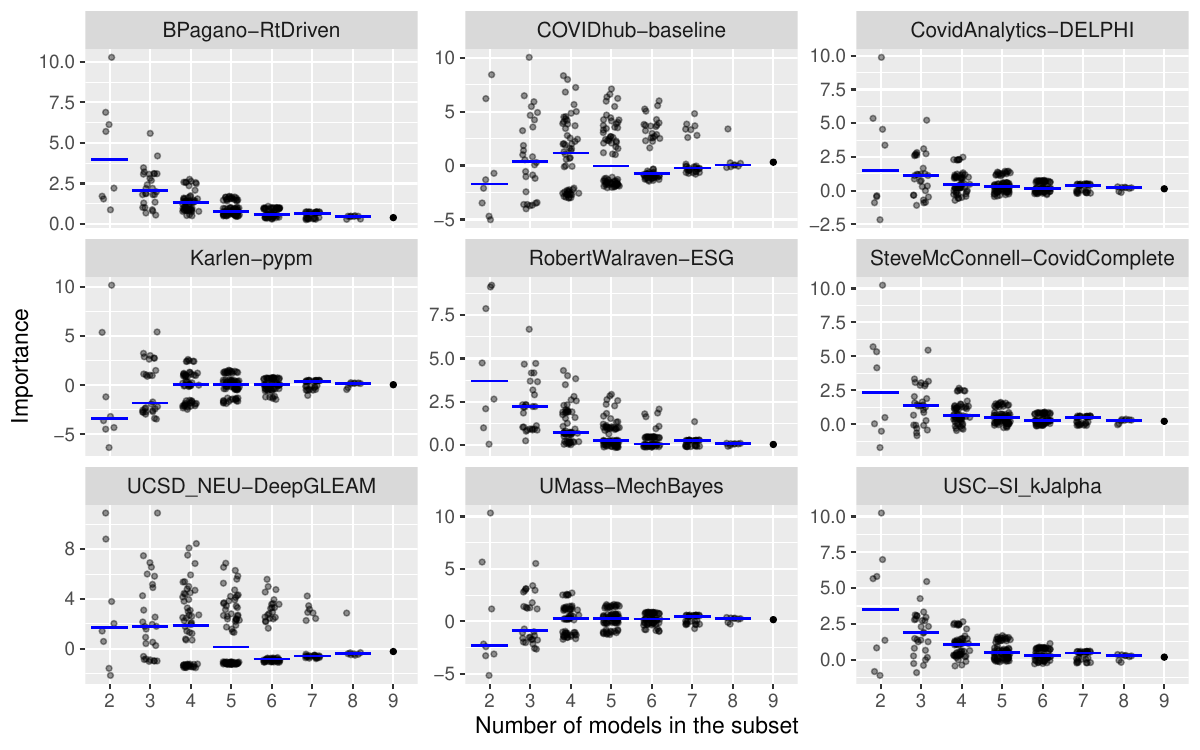}
        \label{fig:imp_by_subset_a}
    \end{subfigure}
    
    \begin{subfigure}[b]{\textwidth}
        \captionsetup{position=above} 
        \caption{Model importance metric distribution (Target end date: 2021-08-28)}
        \includegraphics[page=2, width=\textwidth]{figures/imp_across_subsetsize.pdf}
        \label{fig:imp_by_subset_b}
    \end{subfigure}
    \vspace{-6ex}
    \caption{Model importance metric distribution across different subset sizes for two target end dates with (a) lowest and (b) highest variances of model importance across the nine models over the 53 forecast weeks in 2021, for 4-week ahead incident deaths in Massachusetts. Within each facet, each dot represents the model's importance metric for a subset, and the blue bars indicate the mean value of them across subsets of the corresponding size ($x$-axis). The metrics show high variance in small subsets  (especially in panel (b)) but remain generally stable when only one or two models are removed from the pool of models considered.}
    \label{fig:imp_by_subset}
\end{figure}

\begin{figure}[H]
    \centering
    \begin{subfigure}[b]{\textwidth}
        \captionsetup{position=above} 
        \caption{Forecasts of incident weekly deaths in MA as of 2021-06-26}
        \includegraphics[page=1, width=\textwidth]{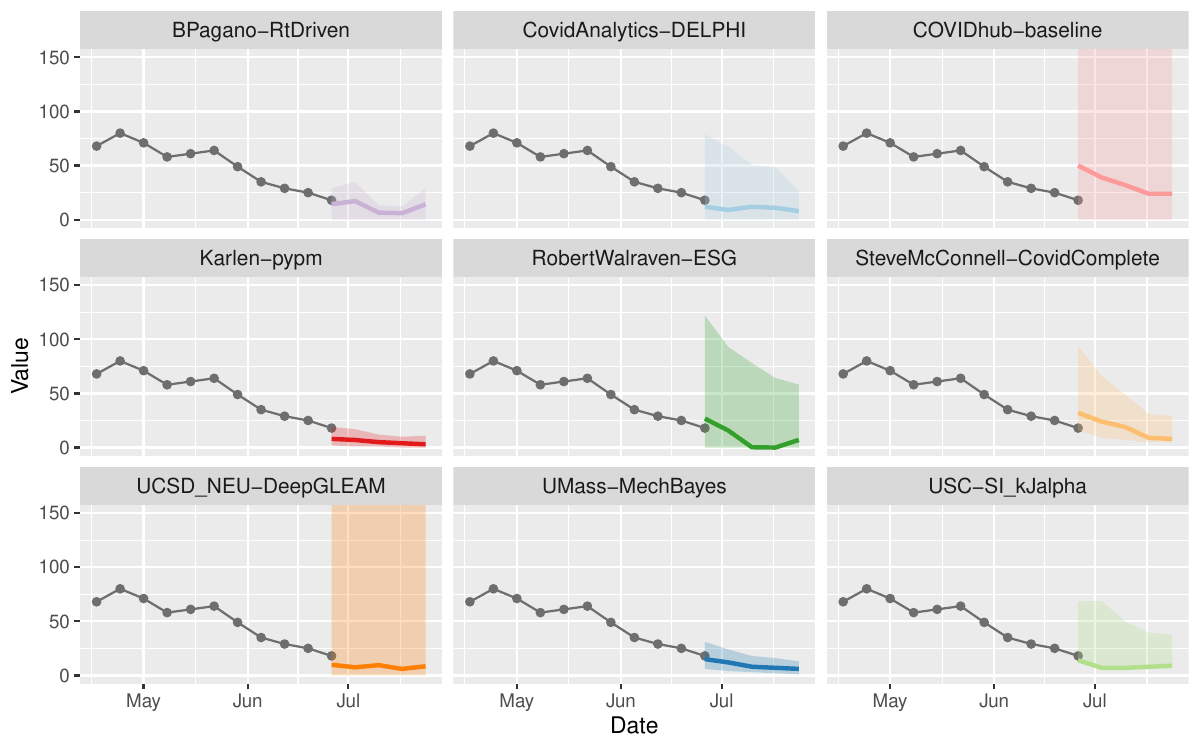}
        \label{fig:forecasts-variation1}
    \end{subfigure}
    \begin{subfigure}[b]{\textwidth}
        \captionsetup{position=above} 
        \caption{Forecasts of incident weekly deaths in MA as of 2021-07-31}
        \includegraphics[page=2, width=\textwidth]{figures/forecasts-variation.pdf}
        \label{fig:forecasts-variation2}
    \end{subfigure}
    \vspace{-6ex}
    \caption{Nine models' 1- through 4-week ahead forecasts of incident weekly deaths in Massachusetts with point predictions and prediction intervals (truncated here for better visibility of the observed data). The gray lines indicate the observed ground truth available at the time the forecasts were generated. The 4-week ahead predictions correspond to the target end dates (a) ``2021-07-24,” illustrating similar model performance, and (b) “2021-08-28,” showing substantial differences across models which lead to large differences in model importance (\hyperref[fig:imp_by_subset_b]{Figure \ref{fig:imp_by_subset_b}}).}
    \label{fig:forecasts-variation_both}
\end{figure}

Across the 53 forecast weeks, the lowest and highest variances of the model importance metrics across the nine participant models were observed for the target end dates ``2021-07-24" and ``2021-08-28", respectively, and the corresponding variances for the nine LOMO values on each date are 0.03 and 387.46. 
We selected these two dates as representative of tasks with high and low variance in model importance.
For each target end date, we examined how the variation in the LOMO metric depends on the subset size and the interaction with other models. For the task with the smallest variance, the scores show little difference across subset size (\hyperref[fig:imp_by_subset_a]{Figure \ref{fig:imp_by_subset_a}}). In contrast, for the task with the largest variance, the importance scores of some models (e.g., Karlen-pypm, UMass-MechBayes, USC-SI\_kJalpha) are strongly affected by the subset size (\hyperref[fig:imp_by_subset_b]{Figure \ref{fig:imp_by_subset_b}}). 
This observation suggests that the importance metric remains relatively stable under different ensemble compositions when all models show similar performance levels (\hyperref[fig:forecasts-variation1]{Figure \ref{fig:forecasts-variation1}}), but becomes sensitive to interaction effects among component models, particularly when there are few models being considered and the performance of certain models differs substantially from that of others (\hyperref[fig:forecasts-variation2]{Figure \ref{fig:forecasts-variation2}}).

Overall, the importance scores of a model tend to exhibit high variability in smaller subsets. LASOMO calculates the importance of model $i$ ($\phi_i^{\text{lasomo}}$) by averaging its mean importance scores across all subset sizes with equal weights:
$$\phi_i^{\text{lasomo}} = \frac{1}{n-1}\sum_{r=2}^{n} \bar{\phi}_{i,r},$$
where $n$ denotes the total number of models in the full set, $r$ denotes the size of subsets that include model $i$, and 
$\bar{\phi}_{i,r}$ represents the mean importance score of the model $i$ across all subsets of size $r$. Thus, the importance metric measured by LASOMO captures the average contribution of a model across different ensemble configurations, but it can also be partially distorted by the subset sizes with high variability. On the other hand, the LOMO model importance scores for a particular model are generally stable when only one or two other models are removed from the pool of models considered.

Our metric is on the same scale as the data, which enhances the interpretability and allows for a more direct comparison of component models' contributions. However, this advantage comes at the cost of the potential instability, which may limit the interpretation of average scores. This is an accepted trade-off of evaluating contributions on the scale of the observed data. We note that some have suggested scoring data on the log scale to reduce the influence of outliers (\citet{bosse2023scoring}) and the importance could be calculated on log-scale data as well. Others have observed that scoring on the original scale can provide meaningful insights about model performance (\citet{Bracher2021}).

\bibliographystyle{model5-names}
\bibliography{references}